\documentclass[final,12pt,authoryear]{elsarticle}
\usepackage{hyperref}
\usepackage{bm}
\usepackage{color}
\usepackage[tbtags]{amsmath}
\usepackage{amssymb,amsthm}

\usepackage{pdfpages}
\usepackage{bbm,endnotes}
\usepackage{mathrsfs} 
\usepackage{pgfplots}
\usepackage{tikz-3dplot}
\usepackage[utf8]{inputenc}
\usepackage[T1]{fontenc}
\usetikzlibrary{arrows.meta,angles,quotes,calc}
\usetikzlibrary{positioning}
\usetikzlibrary{calc}
\usepackage{float} 
\usepackage[textsize=footnotesize]{todonotes}
\usepackage{showlabels}
\newcommand\mynote[1]{\footnote{#1}}

\renewcommand\mynote[1]{}
\newcommand\OK{\color{green}\checkmark\color{black}}
\renewcommand\OK{}

\newcommand\mathone[1]{\overset{\mbox{\tiny 1}}{#1}}
\renewcommand\aa{\boldsymbol a}
\newcommand\yy{\mathbf y}

\newcommand\ba{\boldsymbol a}
\newcommand\bd{\boldsymbol d}
\newcommand\bb{\boldsymbol b}
\newcommand\bc{\boldsymbol c}
\newcommand\be{\boldsymbol e}
\newcommand\bt{\boldsymbol t}
\newcommand\oo{\mathbf o}
\newcommand\bn{\boldsymbol n}

\newcommand\bphi{\boldsymbol\phi}
\newcommand\bu{\boldsymbol u}
\newcommand\bv{\boldsymbol v}
\newcommand\bw{\boldsymbol w}

\newcommand\vv{\boldsymbol v}
\newcommand\bl{\boldsymbol l}
\newcommand\bfx{\mathbf x}
\newcommand\bfy{\mathbf y}
\newcommand\bfo{\mathbf o}

\renewcommand\AA{\boldsymbol A}
\newcommand\bA{\boldsymbol A}
\newcommand\bF{\boldsymbol F}
\newcommand\bG{\boldsymbol G}

\newcommand\bI{\boldsymbol I}
\newcommand\bL{\boldsymbol L}
\newcommand\bM{\boldsymbol M}
\newcommand\bN{\boldsymbol N}
\newcommand\bP{\boldsymbol P}
\newcommand\bS{\boldsymbol S}
\newcommand\bT{\boldsymbol T}
\newcommand\WW{\boldsymbol W}

\newcommand\sa{\boldsymbol{\mathsf a}}
\newcommand\ssb{\boldsymbol{\mathsf b}}
\newcommand\ssc{\boldsymbol{\mathsf c}}
\renewcommand\sl{\boldsymbol{\ell}}
\renewcommand\ss{\boldsymbol{\mathsf s}}
\newcommand\sg{\boldsymbol{\mathsf g}}
\newcommand\sm{\boldsymbol{\mathsf m}}

\newcommand\sq{\boldsymbol{\mathsf q}}
\newcommand\sn{\boldsymbol{\mathsf n}}
\newcommand\st{\boldsymbol{\mathsf t}}
\newcommand\sv{\boldsymbol{\mathsf v}}

\newcommand\sA{\boldsymbol{\mathsf A}}

\newcommand\sE{\boldsymbol{\mathsf E}}
\newcommand\sK{\boldsymbol{\mathsf K}}
\newcommand\sG{\boldsymbol{\mathsf G}}
\newcommand\sN{\boldsymbol{\mathsf N}}
\newcommand\sM{{\boldsymbol{\mathsf M}}}
\newcommand\sP{{\boldsymbol{\mathsf P}}}

\journal{Mechanics of Materials}

\begin{document}
	
	\begin{frontmatter}
		
		\title{\color{black}A coordinate-free guide to the mechanics of thin shells\color{black}}
		
		\author[1]{Giuseppe Tomassetti}
		\ead{giuseppe.tomassetti@uniroma3.it}
		
		\address[1]{Department of Industrial, Electronic, and Mechanical Engineering  \\Roma Tre University\\ Via della Vasca Navale 79, 00146, Rome, Italy}

	\begin{abstract}
		\color{black}In this tutorial, \color{black}we provide a coordinate-free derivation of the system of equations that govern equilibrium of a thin shell that can undergo shear. This system involves tensorial fields representing the internal force and couple per unit length that adjacent parts of the shell exchange at their common boundary.
		By an appropriate decomposition of \color{black}those \color{black}quantities, \color{black}we obtain a representation of the internal power in terms of time derivatives \color{black}of suitable strain measures\color{black}. Subsequently, we propose constitutive equations that employ these strain measures as independent variables\color{black}. \color{black}After specializing \color{black}the theory to the case of unshearable shells, we linearize the resulting equations. As an application, 
		we study the free vibrations of a pressurized spherical shell,  showcasing the advantages of a coordinate-free perspective, which simplifies \color{black}both the deduction and the solution \color{black}of the \color{black}final \color{black}governing equations. 
			\end{abstract}
		
		
	\end{frontmatter}

\section{Introduction}

\subsection{Aim of this tutorial}
Most of the scholarly work dealing with shells makes extensive use of 
coordinate systems, and sometimes of concepts from differential geometry, such as covariant derivatives, Christhoffel symbols, and so on.  The same is true for most textbooks on shell theory \citep{adriaenssensShellStructuresArchitecture2014, ciarletMathematicalElasticity2021,mansfieldBendingStretchingPlates1989a,mollmannIntroductionTheoryThin1981}. In some cases, the treatment is adapted to the special geometry of the shell, and to special loading \color{black}conditions \color{black} \cite{calladineTheoryShellStructures1983}. 

While coordinates undeniably serve a crucial role in solving specific problems, their utilisation may \color{black} have \color{black}the unintended consequence of impeding mechanical insight and obscuring the theory, particularly for those who are new to the field. By relying on coordinates, focus may be shifted towards heavy algebraic manipulations, \color{black}diverting attention from the fundamental mechanical principles that underlie the theory.\color{black} 

The goal of \color{black}the present tutorial \color{black}is to provide a coordinate-free  derivation of the equations that govern the equilibrium of elastic shells using a direct approach. \color{black}We specifically focus on a director-based shell theory characterized by a single director with unit magnitude.  \color{black}Our aim is to demonstrate that a basic understanding of the theory of thin shells can be achieved without relying heavily on differential geometry. In fact, the only concepts we shall need are: a notion of \color{black}the \color{black}superficial gradient; a notion of \color{black}the superficial \color{black}divergence and a version of the \color{black}superficial \color{black}divergence theorem for tangential vector fields; the basic result that the superficial gradient of the \color{black}unit \color{black}normal is \color{black}a symmetric tensor whose range is contained in the tangent space.\color{black}

 In our derivation, we have taken inspiration from \cite{silhavyDirectApproachNonlinear2013}. In particular, we borrow from \cite{silhavyDirectApproachNonlinear2013} the idea of identifying superficial vectors and tensors with linear mappings having their domain of definition in the entire space, rather than the tangent space to the surface. This approach, \color{black}which can be found in several antecedents, with one of the earliest being in the work of \cite{gurtinNatureConfigurationalForces1995}, \color{black}makes it conceptually easier to compare vectors or tensors at two different points of the surface. In contrast to \cite{silhavyDirectApproachNonlinear2013}, however, we consider also shearable shells and, more importantly, our approach is not variational. 
 
Another coordinate-free treatment of two-dimensional material surfaces can be found in \color{black}the work of \color{black}\cite{gurtinContinuumTheoryElastic1975}, where bodies with material boundaries capable of sustaining tension were considered. An extension of  the model to incorporate couple stresses by means of a theory of grade two has been proposed by \cite{murdochDirectNotationSurfaces1978}. \color{black}A coordinate-free \color{black}form of the equilibrium equations, obtained by a deductive approach, are also offered by \cite{favataNewCNTorientedShell2012}. Equations of equilibrium in coordinate-free form have been obtained by  \cite{silhavyDirectApproachNonlinear2013} as Euler--Lagrange equations of an energy functional.
 \color{black}Within \color{black}a purely geometrical setting, the equations of compatibility have been given a coordinate-free formulation by \cite{seguinCoordinatefreeCompatibilityConditions2022}. A coordinate-free version of the Koiter's model, deduced from three-dimensional elasticity through the deductive approach, has been obtained by \cite{steigmannAsymptoticFinitestrainThinplate2007}.

\subsection{Deductive and direct approaches}
\color{black}To place our contribution into perspective, a digression 
concerning the conventional approaches to deriving
theories of thin structures is fitting. \color{black}The problem of establishing consistent theories of shells, or other thin structures such as rods or ribbons, has been addressed through two approaches: the \emph{deductive approach} and the \emph{direct approach} \citep{villaggio1997}. 

The deductive approach pursues a theory of thin structures by a process of ``dimension reduction'' that has its starting point in a three-dimensional theory. Typically, this theory is linear or nonlinear elasticity, or a multiphysics theory having elasticity as its backbone. Nevertheless, there exist derivations \color{black}that begin with \color{black}with the theory of viscous fluids \citep{ribeGeneralTheoryDynamics2002}, \color{black}and \color{black}with atomistic models \citep{aziziBendingStretchingBehavior2023, daviniGaussianStiffnessGraphene2017}

The most elementary approaches to dimension reduction \color{black}hinge on an Ansatz concerning the \color{black}particular form \color{black}of \color{black}the displacement, strain, or stress. Such assumptions are often based on intuition or \color{black} on \color{black}comparisons with special cases \color{black}which admit \color{black}exact solutions. Although the \color{black} Ansatz does not \color{black}satisfy the three-dimensional equations in all cases, an averaging process is employed over the \color{black}thickness of body \color{black}to provide approximate descriptions of the stress, strain, and displacement.

\color{black}More systematic and sophisticated methods for developing theories of thin structures 
\color{black}have also been devised: the asymptotic method \citep{ciarletAsymptoticAnalysisLinearly1996a,ciarletAsymptoticAnalysisLinearly1996,steigmann2007,wang2019}, the variational method \citep{anzellottiDimensionReductionVariational1994,freddiNONLINEARTHINWALLEDBEAMS2012,frieseckeTheoremGeometricRigidity2002}, the constraint and the scaling method \citep{daviConstraintScalingMethods1993,lemboMembranalFlexuralEquations1989,podio-guidugliExactDerivationThin1989,podio-guidugliConstraintScalingMethods1990},  among others. These methods not only offer more systematic derivations, but also make it possible to derive theories of non-conventional thin structures such as ribbons \citep{freddiCorrectedSadowskyFunctional2016},  or structures made of nematic \color{black} elastomers \color{black}\color{black}\citep{agostinianiDimensionReductionGamma2017,davoliMagnetoelasticThinFilms2021}, \color{black}magneto-acive materials \cite{davoliMagnetoelasticThinFilms2021}, thermoelastic materials \cite{favataBeamTheoryConsistent2016}, 
piezoelectric materials \citep{nicotraPiezoelectricPlatesChanging1998,vidoli2000}, polymer gels \citep{andriniTheoreticalStudyTransient2021,lucantonioReducedModelsSwellinginduced2012}, and general morphing materials \citep{argentoTargetMetricShell2021}, 
\color{black}for which standard methods fall short due to lack of intuition and the absence of precedent.\color{black}

With the direct approach, \color{black}a thin structure \color{black} is instead regarded from the outset as a \color{black}lower-dimensional \color{black}
\color{black}continuous body.  To resist bending, this \color{black}body \color{black}must support internal couples, a feature which demands that extra kinematics comes into play, through the inclusion of a set of directors or the second gradients  of the deformation in the list of state variables. 
Examples of application of the direct approach can be found in \color{black}the work of \color{black}\cite{antmanProblemsNonlinearElasticity2005}, \cite{Naghdi1973}, and \cite{rubinCosseratTheoriesShells2000}. \color{black}It is worth noting that, for most director-based 
shell theories (\cite{cohenNonlinearTheoryElastic1966,hilgersELASTICSHEETSBENDING1992,koiter1973foundations,sandersNonlinearTheoriesThin1963}) \color{black}the set of relevant directors reduces to the \color{black} unit \color{black}normal to the surface; \color{black}in this case, the strain energy depends on both the first and second gradients of the deformation.\color{black}

\subsection{\color{black}Outline}
\begin{figure}[t]\label{fig:deformation} 
\centering
\begin{tikzpicture}
\coordinate (P) at (-4,0);
\coordinate (Q) at (3,0);
\coordinate (X) at (-4,0.9);
\coordinate (Y) at (3.5,1.5);
\coordinate (S) at (-3,-0.5);
\coordinate (T) at (Q)+(-6,-2);
\node[draw=none,fill=none] at (P){\includegraphics[width=0.3\paperwidth]{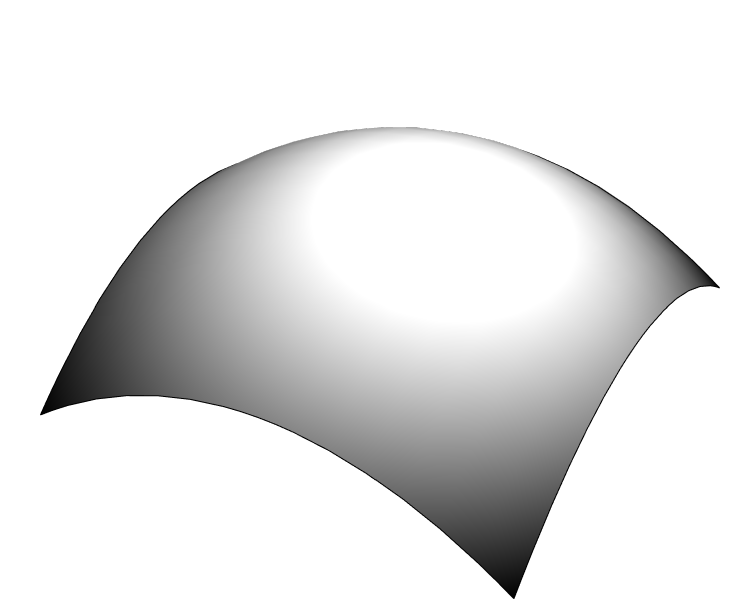}};
\node[draw=none,fill=none] at (Q){\includegraphics[width=0.3\paperwidth]{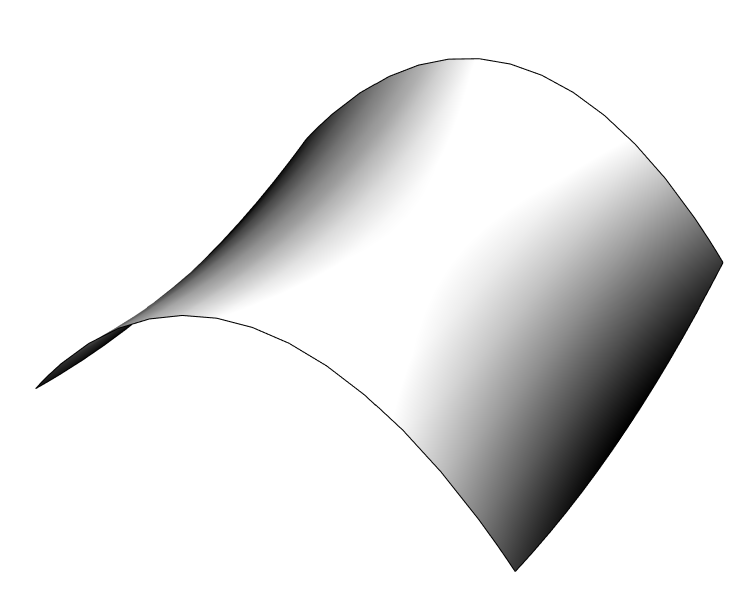}};
\draw [very thick,-stealth] (X) node [left] {$\bfx$}
      to [out=30,in=160] (Y) node [right] {$\bfy(\bfx)$};
\filldraw (X) circle (2pt);
\filldraw (Y) circle (2pt);
\draw  ($(X) + (-0.7,-0.5)$) node [below] {$T_\bfx\mathcal S$} -- ++(1.3,0.3) -- ++(0,0.7) -- 
++(-1.268,-0.3) -- cycle;
\draw  ($(Y) + (-1,-0.5)$) node [below right] {$T_{\bfy(\bfx)}\overline{\mathcal S}$} -- ++(1.8,0.0) -- ++(0.4,0.9) -- 
++(-1.8,-0) -- cycle;
\draw [-stealth,thick] (X) -- +(-0.25,1) node [above] {$\bn(\bfx)$};
\draw [-stealth,thick] (Y) -- +(-0.8,0.7) node [above] {$\bd(\bfx)$};
\draw [-stealth,thick] (Y) -- +(-0.3,0.9) node [right] {$\overline\bn(\bfy(\bfx))$};
\draw (S) node [above] {$\mathcal S$};
\draw (4,-0.5) node [above] {$\overline{\mathcal S}$};
\end{tikzpicture}
\caption{Left: the \color{black}surface \color{black}$\mathcal S$, a typical point $\bfx$, the normal $\bn(\bfx)$, and the tangent space $T_\bfx\mathcal S$ at $\bfx$. Right: the \color{black}image  $\overline{\mathcal S}$ of $\mathcal S$\color{black}, the image $\bfy(\bfx)$ of $\bfx$, the corresponding normal $\overline\bn(\bfy(\bfx))$, and the tangent space $T_{\bfy(\bfx)}\overline{\mathcal S}$ at $\bfy(\bfx)$. \color{black}For an unshearable shell, the director coincides with the current normal, \emph{i.e.},  $\bd(\bfx)=\overline\bn(\bfx)$\color{black}. }
\end{figure}

\color{black}Our presentation is based on the direct approach described above. \color{black}We model a shell as a two dimensional continuous body of oriented particles. We identify the set of such particles with a smooth surface $\mathcal S$ in 
 three-dimensional Euclidean point space $\mathcal E$, and we describe the typical configuration of the shell by assigning to each particle $\bfx\in\mathcal S$ a place $\bfy(\bfx)$ on the deformed surface $\overline{\mathcal S}$, and an orientation, specified by a unit vector $\bd(\bfx)$ as illustrated in Fig. \ref{fig:deformation}.

Our first result is a coordinate-free form of the equilibrium equations that govern the shell:
 \begin{equation}\label{eq:1d}
 \begin{aligned}
& \operatorname{div} \boldsymbol{S}+\boldsymbol{b}=\mathbf{0}, \\
& \operatorname{skw}\left((\operatorname{div} \boldsymbol{M}+\boldsymbol{c}) \otimes \boldsymbol{d}+\boldsymbol{S} \boldsymbol{F}^{\top}+\boldsymbol{M} \boldsymbol{G}^{\top}\right)=\mathbf{0}.
\end{aligned}	
 \end{equation}
 These equations are obtained through an argument based on the invariance of the external power expended on an arbitrary \color{black}subsurface $\mathcal P$ of the reference configuration $\mathcal S$\color{black}.

 Here $\bF=\nabla\bfy$ and $\bG=\nabla\bd$ are the \color{black}(superficial) \color{black}gradients of, respectively, the placement and the director field. Moreover, $\bS$ and $\bM$ are tensor fields on $\mathcal S$ which convey the densities, per unit \emph{referential length}, of force and couple that \color{black}adjacent \color{black}body parts exchange \color{black}across \color{black} their common boundary \emph{in the deformed configuration}. In the theory of shells, the tensors $\bS$ and $\bM$ play \color{black}analogous roles to \color{black}the \emph{Piola stress} in three-dimensional Cauchy continua. Likewise, $\bb$ and $\bc$ are the surface densit\color{black}ies \color{black}of \color{black}the \color{black}body force and \color{black}the \color{black}body couple, \color{black}possibily including inertia\color{black}. 

The tensor fields $\bS$, $\bM$, \color{black}$\bF$, and $\bG$ \color{black}are \emph{superficial}: they are supposed to act solely on vectors on $T_{\bfx}\mathcal S$, the tangent space at $\bfx$. For convenience, however, we extend their domain of definition to $T\mathcal E$, the space of all vectors, by \color{black}stipulating that \color{black}$\bS\bn=\mathbf 0$, $\bM\bn=\mathbf 0$\color{black}, \color{black}and so on. Thus, $\bS$, $\bM$, \color{black}$\bF$,  and $\bG$ \color{black}are linear transformation from $T\mathcal E$, the space of translation\color{black}s \color{black}of $\mathcal E$ into itself, whose null space includes $\bn$. This makes it conceptually easier to compare the values of tensor fields at different point of $\mathcal S$ without introducing \color{black}the differential-geometric notion of \color{black}a connection \citep{lee2003introduction}. 

In the first of \eqref{eq:1d}, the divergence of the superficial tensor field $\bS$ is defined as follows: first, for $\sv$ a tangential vector field, we set ${\rm div}\sv={\rm tr}\nabla\sv=\sP\cdot\nabla\sv$, where $\nabla\sv$ is the surface gradient of $\sv$ and $\sP(\bfx)=\bI-\bn(\bfx)\otimes\bn(\bfx)$ is the projection on the tangent space $T_\bfx\mathcal S$. Then, ${\rm div}\bS$, is the unique vector field such that $\ba\cdot{\rm div}\bS={\rm div}(\bS^\top\ba)$ for every constant vector $\ba$ (note that $\bS^\top\ba$ is a tangential vector field). The divergence of all other tensor fields is defined in a similar fashion. 

In the second of \eqref{eq:1d}, the symbol $\otimes$ denotes the dyadic product: \color{black}given \color{black}two vectors, say $\bu$ and $\bv$, their dyadic product is the tensor defined by $(\bu\otimes\bv)\bw=(\bv\cdot\bw)\bu$ for every vector $\bw$; moreover, the operator $\operatorname{skw}$ applied to a tensor gives its skew-symmetric part. Since the dimension of the space of skew-symmetric tensors is three, the second of \eqref{eq:1d} is equivalent to three-scalar equations; of these equations, the first two can be gathered into 
\begin{equation}\label{eq:2bb}
\bd\times({\rm div}\bM+\bc-\bl)=\mathbf 0,
\end{equation}
where $\bl=(\bS\bF^\top{-}\bF\bS^\top)\bd$ (note that the projection of \eqref{eq:2bb} along $\bd$ vanishes identically). The remaining scalar equation is an algebraic symmetry condition, whose expression is better rendered by introducing the decompositions:
\begin{equation}\label{eq:3ff}
\bS=\bF\sN+\bd\otimes\sq,\qquad \bM=\bF\sM+\bd\otimes\sm,\qquad \bG=\bF\sG+\bd\otimes\sg,
\end{equation}
where $\sN$, \color{black}$\sM$, and \color{black}$\sG$ are \emph{tangential} tensor fields, that is \emph{superficial} tensor fields whose range at a given point $\bfx$ is \emph{orthogonal to }$\bn(\bfx)$ (the referential normal), and $\sq$, \color{black}$\sm$, and \color{black}$\sg$ are tangential vector fields. We refer to $\sN$ and $\sq$ as, respectively, the \emph{membrane-force tensor} and the \emph{shear-force vector}. One may interpret $\sN$ and $\sM$ as the two-dimensional counterparts of the Cosserat stress (also known as second \color{black}Piola--Kirchhoff \color{black}stress). With \eqref{eq:3ff}, the scalar symmetry \color{black}condition \color{black}that complements \eqref{eq:2bb} is
\begin{equation}\label{eq:4hhh}
\color{black}{\rm skw}\tilde\sN=\mathbf 0,\color{black}
\end{equation}
where \color{black}$\tilde{\boldsymbol{\mathsf N}}$, defined by\color{black}
\begin{equation}\label{eq:5hhh}
\color{black}\tilde\sN=\sN-\sG\sM^\top\!\!,	
\end{equation}
 is the \emph{effective membrane-force tensor}.

Our second result, which relies heavily on the decompositions \eqref{eq:3ff} is the \color{black}representation\color{black}
\begin{equation}\label{eq:5bbb}
 W_{\rm int}(\mathcal P)[\dot\bfy,\dot\bd]=\int_{\mathcal P}\left(\tilde\sN\cdot\dot\sE+\sM\cdot\dot\sK+\tilde\sq\cdot\dot\ss\right){\rm d}S
 \end{equation}
\color{black}for the \color{black}internal power \color{black}, \color{black}where $\tilde\sq=\sq-\sM\sg$ is the \emph{effective shear vector}, and 
 \begin{equation}\label{eq:6bb}
 \sE=\frac 12 (\bF^\top\bF-\sP),\qquad \sK=\bF^\top\bG-\nabla\bn,\qquad \ss=\bF^\top\bd,
 \end{equation}
 account, respectively, \color{black}for in \color{black}plane strain, bending, and shearing. The tensor fields $\sE$, $\sK$ and the vector field \color{black}$\sg$ \color{black}vanish when the shell is in the reference configuration, and are invariant under a change of observer, thus they may be considered as appropriate measures of strain. We remark that the expressions \eqref{eq:6bb} are not postulated, but rather follow from the decompositions \eqref{eq:3ff} by working out the expression of the internal power. 
 
  The expression \color{black}\eqref{eq:5bbb} \color{black}of the internal power suggests that constitutive prescriptions should be provided for $\tilde\sN$, $\sM$, and $\tilde\sq$, and that these quantities should depend on $\sE$, \color{black}$\sK$, and $\ss$\color{black}. \color{black}Evidently, \color{black}these prescription should be consistent with the symmetry requirement \eqref{eq:4hhh}.
  To exploit these constitutive equations, the equations of equilibrium $\eqref{eq:1d}_1$ and \color{black}\eqref{eq:2bb} \color{black}should be rewritten as:
\begin{equation}
\label{eq:1b}
\begin{aligned}
&{\rm div}(\bF\sN)+({\rm div}\sq)\bd+\bG\sq+\bb=\mathbf 0,\\
&\bd\times({\rm div}(\bF\sM)-\bF\tilde{\sq}+\bc)=\mathbf 0.
\end{aligned}
\end{equation}
\color{black}The \color{black}substitution of the expressions of the strain measures \eqref{eq:6bb} into the constitutive equations \color{black}then \color{black}leads to a system of \color{black}governing equations \color{black}for the unknowns $\bfy$ and $\bd$.

We next turn our attention to unshearable shells, characterized by the internal constraint:
\begin{equation}\bF^\top\bd=\mathsf s\equiv \mathbf 0.\end{equation}
For unshearable shells, we have
$\sm=\mathbf 0$ \color{black}and \color{black}$\sg=\mathbf 0$, so that $\bF\sM=\bM$ and $\widetilde\sq=\sq$. Furthermore, the bending tensor $\sK$ can be written as
\begin{equation}\label{eq:9}
\sK=-\overline\bn\nabla\bF\color{black}-\nabla\bn,
\end{equation}
where $\overline\bn\nabla\bF$ denotes the \emph{tangential} tensor \color{black}defined such \color{black}that $(\overline\bn\nabla\bF)\cdot\boldsymbol{\mathsf A}=\nabla\bF\cdot(\overline\bn\otimes\boldsymbol{\mathsf A})$ for every second-order superficial tensor $\boldsymbol{\mathsf A}$\color{black}. Thus, unshearable shells are \color{black} in \color{black}all respects bodies of \color{black}second-gradient\color{black}.

For an unshearable shell, the shear vector $\sq$ is a reactive term \color{black}that can be eliminated from the equilibrium equations, leading to \color{black}the system\color{black}
\begin{equation}\label{eq:4b}
\begin{aligned}	
	&\overline\bP{\rm div}\bN+\bG\bF^{-1}({\rm div}\bM+\mathbf c)+\overline\bP\bb=\mathbf 0,\\
	&{\rm div}(\bF^{-1}({\rm div}\bM+\mathbf c))-\bG\cdot\bN+\overline\bn\cdot\bb=0,
\end{aligned}
\end{equation}
where $\overline\bP=\bI-\overline\bn\otimes\overline\bn$ is the orthogonal projection on the space perpendicular to $\overline\bn$,
\begin{equation}
\bN=\bF\sN=\overline\bP\bS,\color{black}	
\end{equation}
and $\bF^{-1}$ is the pseudo-inverse of $\bF$, \color{black} that is, given a point $\bfx$ in $\mathcal S$, $\bF^{-1}(\bfx)$ is \color{black}the inverse of $\bF(\bfx)$ when its domain and its codomain are restricted to $T_\bfx\mathcal S$ and $T_{\bfy(\bfx)}\overline{\mathcal S}$\color{black}, respectively\color{black}.

In particular, if the reference configuration coincides with the \color{black}deformed \color{black}configuration (this choice is possible, in principle, even if the \color{black}deformed \color{black}configuration is not known), then $\bF=\bF^{-1}=\sP$ (we recall that $\sP=\bI-\bn\otimes\bn$ is the orthogonal projector on the tangent \color{black}space \color{black}in the reference configuration), and the above equations reduce to 
\begin{equation}\label{eq:59gg}
\begin{aligned}
	&\sP{\rm div}\bN+\nabla\bn({\rm div}\bM+\bc)+\sP\mathbf b=\mathbf 0,\\
	&{\rm div}(\sP{\rm div}\bM+\bc)-\nabla\bn\cdot\bN+\bn\cdot\mathbf b=0.
\end{aligned}	
\end{equation}
\color{black}The system \color{black}\eqref{eq:59gg} can also be obtained by a formal 
linearization of the equilibrium equations \eqref{eq:4b} for small departures from the reference configuration, \color{black}granted that \color{black}the reference configuration is stress-free. For a transversely-isotropic shell, the linear constitutive equations \color{black}read\color{black}:
\begin{equation}\label{eq:6bc}
\begin{aligned}
&\bN=h\Big(2\mu\boldsymbol\varepsilon+\frac{2\mu\lambda}{2\mu+\lambda}({\rm tr}\boldsymbol\varepsilon)\sP\Big)+\frac{h^3}{12}\nabla\bn\Big(2\mu\boldsymbol\kappa+\frac{2\mu\lambda}{2\mu+\lambda}({\rm tr}\boldsymbol\kappa)\sP\Big),\\
&\bM=\frac{h^3}{12}\Big(2\mu\boldsymbol\kappa+\frac{2\mu\lambda}{2\mu+\lambda}({\rm tr}\boldsymbol\kappa)\sP\Big).
\end{aligned}
\end{equation}
Here $h$ is the thickness of the shell, $\mu$ and $\lambda$ are the Lam\'e moduli of the material that comprises the shell, and 
\begin{equation}\label{eq:7f}
\begin{aligned}
&\boldsymbol\varepsilon=\frac 12(\sP\nabla\sv+\nabla\sv^\top\sP)+w\nabla\bn,\\
&\boldsymbol\kappa=-\sP\nabla\nabla w+\nabla\bn\nabla\sv+\nabla\sv^\top\nabla\bn+w(\nabla\bn)^2+\sP\,\sv\nabla\nabla\bn\color{black},
	\end{aligned}
\end{equation}
are the linear strain and bending tensors, with $w$ and $\sv$, \color{black}respectively, \color{black}the normal and the tangential component of the displacement. Here, consistent with the notation introduced in \eqref{eq:9}, we denote by $\sv\nabla\nabla\bn$ the superficial tensor such that $\sv\nabla\nabla\bn\cdot \sA=\nabla\nabla\sv\cdot(\sv\otimes\sA)$ \color{black} for every second-order superficial tensor $\sA$. Note that $\nabla\bn=\nabla\bn^\top$ and\color{black}, thus, that \color{black}$(\nabla\bn)\bn=\mathbf 0$, \color{black}whence it follows \color{black}that $\nabla\bn$ is a symmetric superficial tensor field. As a consequence, $\boldsymbol\varepsilon$ and $\boldsymbol \kappa$ are symmetric superficial tensor fields. \color{black}If \color{black}the shell is stressed in its reference configuration, the linearization of the equilibrium equations and of the constitutive equations  \color{black}gives rise to \color{black}additional terms, both in the equilibrium equations \eqref{eq:59gg} and in the constitutive equations \eqref{eq:6bc}. \color{black}The explicit form of these terms is also determined.

To illustrate the benefits of a coordinate-free approach, we specifically carry out the linearization of the governing equations for an internally pressurized spherical shell. \color{black}As a result, thanks \color{black}to the coordinate-free approach, \color{black} we obtain the motion equations \color{black}of the shell \color{black}in terms of well-recognizable and familiar differential operators, such as gradient, divergence, and Laplacian, applied to the normal and tangential components of the displacement field. \color{black}As a further result, we derive the characteristic equations that determine the natural frequencies and mode shapes. This task is greatly simplified by the notably simple form 
that the aforementioned differential operators take when applied to spherical harmonics.\color{black}

\section{Thin-shells - preliminaries}\label{sec:preliminaries}
We use the symbol $\mathcal E$ to denote the \color{black}ambient three-dimensional Euclidean point space. \color{black}We denote by $T\mathcal E$ the translation space of $\mathcal E$. We refer to the elements of $\mathcal E$ as \emph{points} and to the elements of $T\mathcal E$ as \emph{vectors}. We \color{black}use the term \color{black}\emph{second-order tensor} \color{black}to refer to a \color{black}linear transformation from $T\mathcal E$ to $T\mathcal E$, and we denote the space of such transformations \color{black}by ${\rm Lin}(T\mathcal E,T\mathcal E)$\color{black}. We \color{black} use the term \color{black}\emph{third-order tensor} any linear transformations that maps vectors into tensors or tensors into vectors.
When a third-order tensor $\mathbb A$ \color{black}is treated \color{black}as a linear transformation of vectors into tensors, we represent its action on \color{black}a \color{black}vector $\boldsymbol{\mathsf a}$ \color{black}by $\mathbb A\boldsymbol{\mathsf a}$\color{black}. Conversely, when $\mathbb A$ \color{black}is regarded \color{black}as a linear transformation of tensors into vectors, we denote its action on \color{black}a second-order \color{black}tensor $\boldsymbol{\mathsf A}$ \color{black}by $\mathbb A\boldsymbol{\mathsf A}$\color{black}. We \color{black} use \color{black}a dot \color{black}to denote \color{black}the standard scalar product between vectors \color{black}or  \color{black}second-order tensors. Furthermore, we let  $\ba\mathbb A$ be the unique \color{black} second-order \color{black}tensor such that $\ba\mathbb A\cdot\boldsymbol{\mathsf A}=\mathbb A\cdot\boldsymbol a\otimes\boldsymbol{\mathsf A}\color{black}=\ba\cdot\mathbb A\boldsymbol{\mathsf A}$ \color{black}for every second-order tensor $\boldsymbol{\mathsf A}$\color{black}.

In what follows, $\mathcal S$ is a smooth oriented surface whose positive unit normal field \color{black} is denoted \color{black}by $\bn$. Given a point $\bfx\in\mathcal S$, $T_{\bfx}\mathcal S$ \color{black} is \color{black}the linear space of vectors orthogonal to $\bn(\bfx)$. We shall make frequent use of the orthogonal projector on $T_\bfx\mathcal S$:
\begin{equation}\label{defp}
	\sP(\bfx)=\bI-\bn(\bfx)\otimes\bn(\bfx).
\end{equation}

\color{black}A vector field $\sv$ such that $\sv\cdot\bn=0$ everywhere on $\mathcal S$ is called \emph{tangential}. A \color{black}tensor field $\AA$ on $\mathcal S$ is \emph{superficial} if $\AA\bn=\mathbf 0$ everywhere on $\mathcal S$. \color{black}A \color{black}superficial tensor field $\sA$ \color{black}that satisfies \color{black}$\sA^\top\bn=\mathbf 0$, \emph{i.e.}, \color{black} has range \color{black}perpendicular to $\bn$ everywhere on $\mathcal S$, \color{black}is also called tangential. \color{black}Equivalently, a tangential field $\sA$ is superficial if and only if $\sP\sA=\sA$. We shall use bold slanted fonts for tangential \color{black} tensor fields that are not superficial\color{black}, and bold sans-serif upright fonts for superficial tensor fields.

We refer to Appendix A.1 for the definition of differential operators acting on scalar, vector, and tensor fields on $\mathcal S$. In particular, for the definition of the gradient operator $\nabla$ and of the \color{black}allied \color{black}divergence $\operatorname{div}$ operator, alongside with a version of the divergence theorem for tangential vector and tensor fields on a surface.

The kinematical descriptors of the shell are: a smooth, invertible placement $\yy:\mathcal S\to\mathcal E$, which maps $\mathcal S$ into another smooth surface $\overline{\mathcal S}$; a director field $\bd:\mathcal S\to \mathcal U$, which maps $\mathcal S$ on the unit sphere $\mathcal U\subset T\mathcal E$. 

We require that $\nabla\bfy(\bfx)+\bd(\bfx)\otimes\bn(\bfx)$ be invertible for all $\bfx\in\mathcal S$. This requirement guarantees that $\nabla\bfy(\bfx)$ is invertible  when its domain and codomain are restricted, respectively, to $T_{\bfx}\mathcal S$ and $T_{\bfx(\bfy)}\overline{\mathcal S}$, and that the director $\bd(\bfx)$ does not belong to $T_{\bfy(\bfx)}\overline{\mathcal S}$, the tangent space of $\overline{\mathcal S}$ at the point $\bfy(\bfx)$. In the reference configuration the deformation is $\bfy(\bfx)=\bfx$ and the director field is given by $\bd(\bfx)=\bn(\bfx)$ for all $\bfx\in\mathcal S$. Granted the smoothness of $\bfy(\cdot)$, the surface $\overline{\mathcal S}$ admits a unit normal everywhere, \color{black}as \color{black}given by
\begin{equation}\label{eq:1e}
\overline\bn=\frac{(\nabla\bfy)^\star\bn}{|(\nabla\bfy)^\star\bn|},	
\end{equation}
with $\bA^\star$, the cofactor of $\bA$, being the unique tensor such that $\bA\bv\times\bA\bw=\bA^\star(\bv\times\bw)$ for all pairs of vectors $\bv$ and $\bw$.

\section{Part-wise equilibrium equations.}
We assume that there exist two superficial tensor fields $\bS$ and $\bM$ such that the external power expended \color{black}on an open subset $\mathcal P$ of the surface $\mathcal S$ \color{black}is \color{black}given by \color{black}
\begin{equation}\label{eq:2b}
  W_{\rm ext}(\mathcal P)[\dot{\bfy},\dot\bd]=\int_{\partial\mathcal P}(\bS\boldsymbol{\mathsf n}_{\mathcal P}\cdot\dot\yy+\bM\boldsymbol{\mathsf n}_{\mathcal P}\cdot\dot{\bd})+\int_{\mathcal P}(\bb\cdot\dot\yy+\bc\cdot\dot\bd),
\end{equation}
where $\boldsymbol{\mathsf n}_{\mathcal P}$ is the tangential field on $\partial\mathcal P$ that has unit magnitude, is orthogonal to $\partial\mathcal P$, and  points away from $\mathcal P$. The vector fields $\bS\boldsymbol{\mathsf n}_{\mathcal P}$ and $\bM\boldsymbol{\mathsf n}_{\mathcal P}$ are the line densities of, respectively \emph{contact forces} and  \emph{contact couples} acting on $\mathcal P$ through $\partial\mathcal P$. We refer to $\bb$ and $\bc$, respectively, as the \color{black} referential densities of \color{black} the \emph{body force} and the \emph{body couple}. 

Since $\bd$ is a unit vector, \color{black} it is orthogonal to $\dot\bd$. \color{black} Thus the term $\bM\sn_{\mathcal P}\cdot\dot\bd$ in the expression of the expended power \eqref{eq:2b} depends only on the projection of $\bM\sn_{\mathcal P}$ on the orthogonal complement of $\bd$. We therefore assume, without loss of generality, that the range of $\bM$ is orthogonal to $\bd$, \color{black} namely that\color{black}
\begin{equation}\label{eq:2}
\bM^\top\bd=\boldsymbol{0}.	
\end{equation}
We require that the external power be invariant under the superposition of a rigid velocity field
\begin{equation}
\begin{aligned}
&\dot{\yy}\mapsto\dot{\yy}+\vv+\WW(\yy-\oo),\\
&\dot{\bd}\mapsto\dot{\bd}+\WW\bd,
\end{aligned}
\end{equation}
where $\vv$ is a \color{black}constant \color{black} vector, $\WW$ is a \color{black}constant \color{black}skew-symmetric tensor, and $\mathbf o$ is the point corresponding to some chosen origin; \color{black}this  \color{black}requirement leads to the part-wise balance equations
\begin{equation}\label{eq:1}
\begin{aligned}
&\int_{\partial\mathcal P}\bS\sn+\int_{\mathcal P}\bb=\mathbf 0,\\
&\int_{\partial\mathcal P}(\bS\sn\wedge (\yy-\oo)+\bM\sn\wedge\bd)+\int_{\mathcal P}(\bb\wedge(\yy-\oo)+\bc\wedge\bd)=0,
\end{aligned}
\end{equation}
where $\mathbf a\wedge\mathbf b=\mathbf a\otimes\mathbf b-\mathbf b\otimes\mathbf a$. As shown in Appendix A.2, the use of the divergence theorem yields
\begin{equation}\label{eq:44}
  \begin{aligned}
    &\int_{\mathcal P}\color{black}({\rm div}\bS+\bb)\color{black}=\mathbf 0,\\
    &\int_{\mathcal P}\left(({\rm div}\bM+\bc)\wedge\bd+({\rm div}\bS+\bb)\wedge (\yy-\oo)+2{\rm skw}(\bS\nabla\yy^\top+\bM\nabla\bd^\top)\right)=\mathbf 0,
    \end{aligned}
  \end{equation}
  where, for $\bA$ a tensor, ${\rm skw}\bA=\frac 1 2(\bA-\bA^\top)$ is the skew-symmetric part of \color{black}the second-ordet tensor \color{black}$\bA$.
  
\def\pippo{It is possible to 
start by defining the external power to be
\begin{equation}\label{eq:2bis}
  W_{\rm ext}(\mathcal P)[\dot{\bfy},\dot\bd]=\int_{\partial\mathcal P}(\bt_{\mathcal P}\cdot\dot\yy+\boldsymbol m_{\mathcal P}\cdot\dot{\bd})+\int_{\mathcal P}(\bb\cdot\dot\yy+\bc\cdot\dot\bd),
\end{equation}
where $\bt_{\mathcal P}$ and $\boldsymbol{m}_{\mathcal P}$ are the line densities of the contact forces and contact couples acting on the boundary of the part $\mathcal P$. The existence of  tensorial quantities $\bS$ and $\bM$ that do not depend on $\mathcal P$ and that satisfy $\bS\boldsymbol{\mathsf n}=\bt_{\mathcal P}$ and $\bM\boldsymbol{\mathsf n}_{\mathcal P}=\boldsymbol m_{\mathcal P}$ can be obtained by a construction similar to that used to prove the existence of the Cauchy stress. The procedure is illustrated in the treatise \cite{Naghdi1973} and in the paper \cite{gurtinContinuumTheoryElastic1975}. }

 \section{Point-wise equilibrium equations.}
Henceforth we set
  \begin{equation}\label{eq:12}
    \bF=\nabla\yy,\qquad\bG=\nabla\bd.
  \end{equation}
  We note that, since $\bd\cdot\bd=1$, $\nabla \bd^\top\bd=\boldsymbol 0$, that is, 
\begin{equation}\label{eq:16hhh}
\bG^\top\bd=\mathbf 0.	
\end{equation}
  Thanks to the arbitrariness of the part $\mathcal P$ in \eqref{eq:44}, a standard localization argument yields the point-wise equilibrium equations \eqref{eq:1d}, which we repeat below for the reader's convenience:
  \begin{subequations}\label{eq:34}
 \begin{align}
      &\operatorname{div} \boldsymbol{S}+\boldsymbol{b}=\mathbf{0},\label{eq:3}\\
      &{\rm skw}\left(({\rm div}\bM+\bc)\otimes\bd+\bS\bF^\top+\bM\bG^\top\right)=\mathbf 0.\label{eq:4}
 \end{align}
  \end{subequations}
  The first of \eqref{eq:34} coincides with (4.9) of \color{black}\cite{dicarloShellsThicknessDistension2001}; \color{black} the second can be obtained by combining \color{black}(5.14) \color{black} with \color{black}(4.23b) of \cite{dicarloShellsThicknessDistension2001}\color{black}.
  
\color{black}Equation \color{black}\eqref{eq:4} is equivalent to \emph{two scalar \color{black}partial-differential \color{black}equations}, and \emph{one scalar algebraic equation}. To verify this assertion, let us \color{black}consider \color{black}any vector \color{black}field \color{black}perpendicular to $\bd$. With \color{black}a \color{black}slight abuse of notation, let us denote such vector by $\dot\bd$ (later, we will use the results obtained here with $\dot\bd$ being the actual rate of change of $\bd$). On taking the scalar product of both sides of \eqref{eq:4} with $2\dot\bd\otimes\bd$ we obtain
\begin{equation}\label{eq:21hhh}
\dot\bd\cdot(\operatorname{div}\bM+\bc)=2\dot\bd\cdot \operatorname{skw}(\bF\bS^\top+\bG\bM^\top)\bd.
\end{equation}
Using \eqref{eq:2} and \eqref{eq:16hhh}, we see that $\bG\bM^\top\bd=\bM\bG^\top\bd=\mathbf 0$, and hence, on setting
\begin{equation}\label{eq:7}
   \bl=2{\rm skw}(\bF\bS^\top)\bd,
\end{equation}
we \color{black} find that \color{black}\eqref{eq:21hhh} \color{black}can be expressed as\color{black}
\begin{equation}
	\dot\bd\cdot(\operatorname{div}\bM+\bc)=\dot\bd\cdot\bl.
\end{equation}
Thus, a first consequence of \eqref{eq:4} is
\begin{equation}\label{eq:6}
  \bd\times({\rm div}\bM+\bc-\bl)=\mathbf 0,
\end{equation}
which is equivalent to two scalar \color{black}partial-differential \color{black}equations. The other consequence of \eqref{eq:4} is obtained by pre-multiplying and post-multiplying \eqref{eq:4} by $\color{black}\bP_{\boldsymbol d}=\bI-\bd\otimes\bd$. This yields, in view of \eqref{eq:2}--\eqref{eq:16hhh},
\begin{equation}\label{eq:6b}
{\rm skw}(\color{black}\bP_{\boldsymbol d}\color{black}\bS\bF^\top\color{black}\bP_{\boldsymbol d}\color{black}+\bM\bG^\top)=\mathbf 0.
\end{equation}
By construction, the left-hand side of \eqref{eq:6b} is a skew-symmetric tensor that\color{black}, at a given point $\bfx$ of $\mathcal S$, \color{black}maps the orthogonal complement of $\bd(\bf x)$ into itself, thus has rank at most one. Therefore, \eqref{eq:6b} is equivalent to one scalar equation.

We will see shortly that \eqref{eq:6b} can be cast in a form that looks more convenient in the light of the constitutive theory that we develop below, \color{black}the \color{black}starting point \color{black}of which \color{black}is an expression of the internal power in terms of appropriate strain tensors. As a preliminary step, we \color{black}next \color{black}introduce suitable decompositions of the tensorial quantities of interest.

\section{Decomposition}
As a preliminary step, we introduce a particular pseudo-inverse 
of  $\bF$. First, we set
\begin{equation}\label{eq:22bbb}
\widetilde\bF=\bF+\bd\otimes\bn.	
\end{equation}
In Section \ref{sec:preliminaries}, we assumed that $\widetilde\bF(\bfx)$ be an invertible tensor. As a consequence of that assumption, we can define \color{black}the inverse ${\bF}^{-1}$ of $\boldsymbol{F}$ by\color{black}
\begin{equation}\label{eq:fminus1}
	\bF^{-1}=\sP\widetilde{\bF}^{-1},
\end{equation}
where we recall from \eqref{defp} that $\sP(\bfx)$ is the orthogonal projector on $T_\bfx\mathcal S$, the tangent space of $\mathcal S$ at $\bfx$.
We observe that $\widetilde\bF^{-1}\bd=\bn$, so that
\begin{equation}\label{eq:32ggg}
	\bF^{-1}\bd=\boldsymbol 0.
\end{equation}
We also observe that $\widetilde\bF$ is a bijection between the orthogonal complement of $\bn$ and the orthogonal complement of $\overline\bn$. Thus, for every vector $\bv$ such that $\bv\cdot\overline\bn=0$, we have $\widetilde \bF^{-1}\bv\cdot\bn=\color{black}\boldsymbol 0$, whence $\bv\cdot\widetilde\bF^{-\top}\!\!\bn=0$. Thus, $\widetilde\bF^{-\top}\!\!\bn$ is parallel to $\overline\bn$ and therefore we can write $\widetilde\bF^{-\top}\!\!\bn=\alpha\overline\bn$ for some $\alpha\in\mathbb R$. Then, by \eqref{eq:32ggg}, we have $\alpha\bd\cdot\overline\bn=\bd\cdot\widetilde\bF^{-\top}\bn=\widetilde\bF^{-1}\bd\cdot\bn=1$. \color{black}Altogether\color{black}, we have
\begin{equation}\label{eq:33bbb}
{\widetilde\bF}^{-\top}\bn=(\bd\cdot\overline\bn)^{-1}\bn.
\end{equation}
Using \eqref{eq:fminus1}--\eqref{eq:32ggg}, we compute $\bF^{-1}\bF=\sP\widetilde\bF^{-1}\bF=\sP\widetilde\bF^{-1}(\widetilde\bF-\bd\otimes\bn)=\sP-\sP\widetilde\bF^{-1}\bd\otimes\bn=\sP-\bF^{-1}\bd\otimes\bn=\sP$. Likewise, using \eqref{eq:33bbb} we write $\bF\bF^{-1}=(\widetilde\bF-\bd\otimes\bn)(\bI-\bn\otimes\bn)\widetilde\bF^{-1}=\bI-\widetilde\bF\bn\otimes\widetilde\bF^{-\top}\!\!\bn=\bI-\bd\otimes(\bd\cdot\overline\bn)^{-1}\overline\bn$. In conclusion, we have established that
\begin{subequations}
	\begin{align}
	&\bF^{-1}\bF=\sP,\label{eq:34ttt}\\
	&\bF\bF^{-1}=\overline\bP_{\bd},\label{eq:34bbb}
\end{align}
\end{subequations}
where
\begin{equation}\label{defpbar}
	\overline\bP_{\bd}=\bI-(\bd\cdot\overline\bn)^{-1}\bd\otimes\overline\bn.
\end{equation} 
We notice that $\overline\bP_{\bd}$ is the identity on the orthogonal complement of $\overline\bn$, and \color{black}that\color{black}
\begin{equation}
	\overline\bP_{\bd}\bd=\mathbf 0.
\end{equation}
\color{black}A\color{black}dditional discussion \color{black}concerning \color{black}$\bF^{-1}$ is contained in Appendix A.3.

The introduction of the pseudo-inverse $\bF^{-1}$ opens the way to a convenient decomposition of the tensorial quantities of interest. As to the tensor field $\bS$, we can write $\bS=\bI\bS=\overline\bP_{\bd}\bS+(\bd\cdot\overline\bn)^{-1}(\bd\otimes\overline\bn)\bS=\bF\bF^{-1}\bS+(\bd\cdot\overline\bn)^{-1}\bd\otimes\bS^\top\overline\bn$. Thus, on setting
\begin{equation}\label{eq:11b}
    \sN=\bF^{-1}\bS\quad\text{and}\quad\sq=(\bd\cdot\overline\bn)^{-1}\bS^\top\overline\bn,
\end{equation}
we have the decomposition
\begin{equation}\label{eq:5a}
	\bS=\bF\sN+\bd\otimes\sq.
\end{equation}
We note that, since the range of $\bF^{-1}$ is orthogonal to $\bn$\color{black}, \color{black}the tensor field $\sN$ is superficial, \emph{i.e.}, \color{black}$\sN$ \color{black}maps tangential tensors into tangential tensors. Likewise, since $\bS$ is tangential (\emph{i.e.} $\bS\bn=\mathbf 0$), the vector field $\sq$ is tangential. We shall refer to $\sN$ and $\sq$, respectively, \color{black}as \color{black}the membrane-force tensor and the shear-force vector.

The argument leading to \eqref{eq:11b} can be repeated to obtain the decompositions
  \begin{equation}\label{eq:5b}
    \begin{aligned}
      \bM=\bF\sM+\bd\otimes\sm,\qquad 
      \bG=\bF\sG+\bd\otimes\sg,\\
      \end{aligned}
\end{equation}
where $\sM$ and $\sG$ are superficial tensor fields and $\sm$ and $\sg$ are tangential tensor fields. Concerning \eqref{eq:5b}, we observe for later use that, since $\bM^\top$ and $\bG^\top$ annihilate $\bd$ (recall \eqref{eq:2} and \eqref{eq:16hhh}), we have
\begin{equation}\label{eq:37uu}
	\sM^\top\!\bF^\top\bd+\sm=\mathbf 0,\qquad \sG^\top\!\bF^\top\bd+\sg=\mathbf 0,
\end{equation}
Using the decompositions \eqref{eq:5a}, \eqref{eq:5b}, and \eqref{eq:5c}, we can rewrite \eqref{eq:3} and \eqref{eq:4}, respectively as
\begin{equation}\label{eq:30}
\begin{aligned}
{{\rm div}(\bF\sN)+({\rm div}\sq)\bd+\bG\sq+\bb=\mathbf 0\color{black},}
\end{aligned}
\end{equation}
and
\urldef\myurl\url{https://uniroma3-my.sharepoint.com/personal/gitomassetti_os_uniroma3_it/_layouts/OneNote.aspx?id=%2Fpersonal%2Fgitomassetti_os_uniroma3_it%2FDocuments%2FNotebooks%2FJournal&wd=target%280%20Year%202022%2FDec%202022.one%7CA7482B2A-331E-4705-9160-6E59866C2EAC%2FCalculation%20of%20the%20formula%20for%20l%7C31D94BA9-093C-5744-8410-C37CFC8D885C%2F%29}
\mynote{\myurl}
       \begin{equation}\label{eq:11}
         {\rm skw}(({\rm div}(\bF\sM)+\bc-\bF(\sq-\sM\sg))\otimes\bd+\bF(\sN-\sG\sM^\top)\bF^\top)=\mathbf 0.
       \end{equation}
       Now, we pre-multiply and post-multiply both sides of \eqref{eq:11} by $\bF^{-1}$ and $\bF^{-\top}$. Since $\bF^{-1}\bd=\color{black}\boldsymbol 0$, 
       we obtain $\bF^{-1}\operatorname{skw}(\bF(\sN-\sG\sM^\top)\bF^\top)\bF^{-\top}=\mathbf 0$, that is, $\operatorname{skw}(\bF^{-1}\bF(\sN-\sG\sM^\top)\bF^\top\bF^{-\top})=\mathbf 0$. Recalling the first of \eqref{eq:34ttt}, and noting that $\sN$, $\sG$, and $\sM$ are superficial tensor fields, we arrive at the following symmetry condition:
       \begin{equation}\label{eq:12}
         {{\rm skw}(\sN-\sG\sM^\top)=\mathbf 0.}
         \end{equation}
         Thus, \eqref{eq:11} reduces to  ${{\rm skw}\big(({\rm div}(\bF\sM)+\bc-\bF(\sq-\sM\sg))\otimes\bd\big)=\mathbf 0,	}$
         which can be re\color{black}cast \color{black}as
         \begin{equation}\label{eq:38}
         \bd\times(\operatorname{div}(\bF\sM)+\bc-\bF(\sq-\sM\sg))=\mathbf 0.	
         \end{equation}
Equations \eqref{eq:12} and \eqref{eq:38} are equivalent to, respectively, \eqref{eq:6b} and \eqref{eq:6}.

\section{Internal power and strain measures}
 \subsection{Internal power} Starting from the definition of external power in \eqref{eq:2}, using the divergence theorem and exploiting the equilibrium equations \eqref{eq:3} and \eqref{eq:6}, we write
 \begin{align}
      W_{\rm ext}(\mathcal P)[\dot{\bfy},\dot\bd]&=\int_{\mathcal P}\left(\bS\cdot\dot\bF+({\rm div}\bS+\bb)\cdot\dot\yy+\bM\cdot\dot\bG+({\rm div}\bM+\bc)\cdot\dot\bd\right)\nonumber\\
      &=\int_{\mathcal P}\color{black}\Big(\color{black}\bS\cdot\dot\bF+\bM\cdot\dot\bG+\bl\cdot\dot\bd\color{black}\Big)\color{black}.\label{eq:39p}
      \end{align}
Motivated by \eqref{eq:39p}, we define, for every part $\mathcal P$, the \emph{internal power} expended on a a pair $(\dot\bfy,\dot\bd)$ with $\bd\cdot\dot\bd=0$ as
 \begin{equation}\label{eq:15b}
 W_{\rm int}(\mathcal P)[\dot{\bfy},\dot\bd]:=\int_{\mathcal P}\left(\bS\cdot\dot\bF+\bM\cdot\dot\bG+\bl\cdot\dot\bd\right).
\end{equation}
The definition \eqref{eq:15b} applies for any assignment of the  tensors field $\bS$, $\bM$, and of the vector field $\bl$, irrespectively of whether these fields satisfy the equilibrium equations \eqref{eq:3} and \eqref{eq:6}. The calculations in \eqref{eq:39p} show, however, that if the equilibrium equations \eqref{eq:3} and \eqref{eq:6} hold, then the external power and the internal power coincide:
\begin{equation}
W_{\rm int}(\mathcal P)[\dot\bfy,\dot\bd]=W_{\rm ext}(\mathcal P)[\dot\bfy,\dot\bd].	
\end{equation}

\subsection{Strain measures} We next exploit the decompositions \eqref{eq:5a} and \eqref{eq:5b} to obtain an alternative expression of the internal power in terms of rates of suitable strain descriptors. We begin with the first term under integral sign on the right-hand side of \eqref{eq:15b}. By \eqref{eq:5a}, we can write
\begin{equation}\label{eq:95d}
\bS\cdot\dot\bF=\bF\sN\cdot\bF+\bd\otimes\sq\cdot\dot\bF=\sN\cdot\bF^\top\dot\bF+\sq\cdot\dot\bF^\top\bd.
\end{equation}
We next turn our attention to the second term under integral sign in \eqref{eq:15b}. Using the decompositions \eqref{eq:5b}, we \color{black}obtain\color{black}
\begin{equation}\label{eq:96d}
\begin{aligned}[b]
	\!\!\!\!\!\bM\cdot\dot{\bG}&=(\bF\sM+\bd\otimes\sm)\cdot\dot{\bG}=\sM\cdot\bF^\top\dot{\bG}+\dot{\bG}^\top\bd\cdot\sm\\
	&=\sM\cdot\dot{\overline{\bF^\top\bG}}-\sM\cdot\dot{\bF}^\top\bG+\dot{\overline{\bG^\top\bd}}\cdot\sm-\bG^\top\dot{\bd}\cdot\sm\\
	&=\sM\cdot\dot{\overline{\bF^\top\bG}}-\bG\sM^\top\cdot\dot{\bF}+\dot{\overline{\bG^\top\bd}}\cdot\sm-\bG\sm\cdot\dot{\bd}\\
	&=\sM\cdot\dot{\overline{\bF^\top\bG}}-\bF\sG\sM^\top\cdot\dot{\bF}-(\bd\otimes\sg)\sM^\top\cdot\dot\bF+\dot{\overline{\bG^\top\bd}}\cdot\sm\\
	&\,\quad-\bF\sG\sm\cdot\dot{\bd}-(\bd\otimes\sg)\sG\sm\cdot\dot\bd\\
	&=\sM\cdot\dot{\overline{\bF^\top\bG}}-\sG\sM^\top\cdot\bF^\top\dot{\bF}-\sM\sg\cdot\dot\bF^\top\bd\\
	&\,\,\quad +\dot{\overline{\bG^\top\bd}}\cdot\sm-\sG\sm\cdot\bF^\top\dot{\bd}\color{black}.\color{black}
\end{aligned}	
\end{equation}
In the last of the above chain of equalities we have used the orthogonality between $\dot\bd$ and $\bd$ to eliminate the last term in the fourth line.

We now consider the last term \color{black}of the integrand \color{black}in \eqref{eq:15b}. Using \eqref{eq:33bbb} and \eqref{defpbar}, we write $\bl=\bI\bl=\overline\bP_{\!\bd\,}\bl+(\bd\cdot\overline\bn)^{-1}(\bd\otimes\overline\bn)\bl=\bF\bF^{-1}\bl+(\boldsymbol{d} \cdot \overline{\boldsymbol{n}})^{-1}(\bl\cdot\overline\bn)\bd$. \color{black}We then \color{black}set $\sl=\bF^{-1}\bl$ and $\lambda=(\bd\cdot\overline\bn)^{-1}\bl\cdot\overline\bn$, \color{black}giving \color{black}
\begin{equation}\label{eq:5c}
	\bl=\bF\sl+\lambda\bd\color{black},\color{black}
\end{equation}
thus, can write
\begin{equation}\label{eq:97d}
\bl\cdot\dot\bd=\bF\sl\cdot\bd+\lambda\bd\cdot\dot\bd=\sl\cdot\bF^\top\dot\bd.	
\end{equation}
\color{black}Combining \color{black}\eqref{eq:95d}, \eqref{eq:96d}, and \eqref{eq:97d}, we obtain 
        \begin{multline}\label{eq:20d}
    W_{\rm int}(\mathcal P)[\dot\bfy,\dot\bd]\\
    \qquad=\int_{\mathcal P}\Big((\sN-\sG\sM^\top){\cdot}\bF^\top\dot{\bF}{+}\sM{\cdot}\dot{\overline{\bF^\top\bG}}{+}(\sq-\sM\sg){\cdot}\dot{\bF}^\top\!\!\bd{+}(\sl-\sG\sm){\cdot}\bF^\top\dot{\bd}\Big).
 \end{multline}
\color{black}We now return \color{black}to the definition of $\bl$ in \eqref{eq:7}, and we use, \color{black}in order\color{black}, the decomposition \eqref{eq:5a} of $\bS$, the symmetry condition \eqref{eq:12}, and the relations \eqref{eq:37uu} to \color{black}obtain\color{black}
\begin{align}
	\bl&=\bF\bS^\top\bd-\bS\bF^\top\bd\nonumber\\
	&=\bF\sN^\top\bF^\top\bd+\bF\sq-\bF\sN\bF^\top\bd-(\bF\sq\cdot\bd)\bd\nonumber\\
	&=\bF\sq-(\bF\sq\cdot\bd)\bd+\bF\sG\sM^\top\bF^\top\bd-\bF\sG\sM^\top\bF^\top\bd\nonumber\\
	&=\bF\sq-(\bF\sq\cdot\bd)\bd+\bF\sG\sm-\bF\sM\sg.\label{eq:55yyy}
\end{align}
Since the decomposition \eqref{eq:5c} is unique, from \eqref{eq:55yyy} we deduce \color{black}that\color{black}
\begin{equation}\label{eq:15}
	\sl=\sq+\sG\sm-\sM\sg.
\end{equation}       
Using \eqref{eq:15}, and bearing in mind that $\sN-\sG\sM^\top$ is symmetric,  we obtain, from \eqref{eq:20d}, the following representation for the internal power:
  \begin{equation}\label{eq:20e}
    W_{\rm int}(\mathcal P)[\dot\bfy,\dot\bd]=\int_{\mathcal P}\Big((\sN-\sG\sM^\top)\cdot\frac 12\dot{\overline{\bF^\top{\bF}}}+\sM\cdot\dot{\overline{\bF^\top\bG}}+(\sq-\sM\sg)\cdot\dot{\overline{{\bF}^\top\bd}}\Big).
  \end{equation}
The representation \eqref{eq:20e} of the internal power \color{black}motivates us to \color{black}set
\begin{equation}\label{eq:21b}
	{\tilde\sN=\sN-\sG\sM^\top,\qquad \tilde\sq=\sq-\sM\sg,}
\end{equation}
and \color{black}to \color{black}introduce the strain measures
\begin{equation}\label{eq:16}
{\sE=\frac 12 (\bF^\top\bF-\sP),\quad \sK=\bF^\top\bG-\nabla\bn,\quad \ss=\bF^\top\bd,	}
\end{equation}
\color{black} where we \color{black}recall that $\sP(\bfx)$ is the orthogonal projection on $T_{\bfx}\mathcal S$. We note on passing that the strain measures vanish when $\bfy(\bfx)=\bfx$ and $\bd(\bfx)=\bn(\bfx)$, \emph{i.e.}, when the shell is in its reference configuration. By making use of \eqref{eq:21b} and \eqref{eq:16}, we can \color{black}express \color{black}\eqref{eq:20e} as
\begin{equation}\label{eq:21bb}
W_{\rm int}(\mathcal P)[\dot\bfy,\dot\bd]=\int_{\mathcal P}\big(\tilde\sN\cdot\dot\sE+\sM\cdot\dot\sK+\tilde\sq\cdot\dot\ss\big).	
\end{equation}
We call $\sE$, $\sK$\color{black}, \color{black}and $\ss$, respectively, the \emph{\color{black}stretching \color{black}tensor}, the \emph{bending tensor}, and the \emph{shear vector}.

\subsection{Dissipation principle} For $\psi$ the free-energy density per unit reference area, the dissipation principle asserts that 
\begin{equation}\label{eq:20}
\frac{{\rm d}}{{\rm d}t}\int_{\mathcal P}\psi\le W_{\rm ext}(\mathcal P)	[\dot\bfy,\dot\bd]=W_{\rm int}(\mathcal P)[\dot\bfy,\dot\bd],
\end{equation}
during every admissible process. The arbitrariness of the part $\mathcal P$ in \eqref{eq:20} yields, by localization, the point-wise version 
\begin{equation}\label{eq:22b}
\dot\psi\le 	\tilde\sN\cdot\dot\sE+\sM\cdot\dot\sK+\tilde\sq\cdot\dot\ss
\end{equation}
of the dissipation principle. Motivated by the form of the right-hand side of \eqref{eq:22b}, we select as state variables the strain measures $\sE$, $\sK$ and $\ss$, and we assume that the free-energy density $\psi$ obeys the constitutive equation:
\begin{equation}
\psi=\hat\psi(\sE,\sK,\ss).	
\end{equation}
Then, on requiring, in the manner introduced \color{black}by \cite{coleman1963thermodynamics}\color{black}, that \eqref{eq:22b} be satisfied for whatever local continuation of any conceivable process, we obtain the constitutive equations:
\begin{equation}\label{eq:23}
    {\begin{aligned}
      &\tilde\sN=\partial_{\sE}\hat\psi(\sE,\sK,\ss),\\&\sM=\partial_{\sK}\hat\psi(\sE,\sK,\ss),\\
      &\tilde\sq=\partial_{\ss}\hat\psi(\sE,\sK,\ss).
      \end{aligned}}
      \end{equation}
The equilibrium equation \eqref{eq:12},  \color{black}namely\color{black}
\begin{equation}
\operatorname{skw}\widetilde\sN=\mathbf 0,
\end{equation}
is identically satisfied. The remaining equilibrium equations, \color{black}namely \color{black}\eqref{eq:30} and \eqref{eq:38}, can now be written in terms of $\widetilde\sN$, $\sM$ and $\widetilde\sq$:
\begin{equation}\label{eq:54o}
\begin{aligned}
&\operatorname{div}(\bF(\widetilde\sN+\sG\sM^\top))+\operatorname{div}(\widetilde\sq+\sM\sg)\bd+\bG (\widetilde\sq+\sM\sg)+\boldsymbol{b}=\mathbf{0},\\
&\bd\times\operatorname{div}(\bF\sM+\bc-\bF\widetilde\sq)=\mathbf 0\color{black};\color{black}
\end{aligned}
\end{equation}
\color{black}\eqref{eq:54o} \color{black}can be cast into a system of partial differential equations where the \color{black}unknown quantities \color{black}are the deformation $\bfy$ 
and of the director field $\bd$. In fact, $\tilde\sN$, $\sM$, and $\ss$ depend, through the constitutive 
equations \eqref{eq:23}, on the strains $\sE$, $\sK$, and $\ss$, which in turn can be expressed in terms of $\bF=\nabla\bfy$, $\bG=\nabla\bd$, and $\bd$. Likewise the superficial tensor field $\sG$ defined in the second equation of \eqref{eq:5b} can be expressed in terms of $\sE$ and $\sK$: \begin{equation}\label{eq:56y}
 	\sG=\bF^{-1}\bG=\bF^{-1}\bF^{-\top}\bF^\top \bG=(\bF^{\top}\bF)^{-1}(\sK+\nabla\bn)=(2\sE+\sP)^{-1}(\sK+\nabla\bn).
 \end{equation}
Finally, $\sg$ is determined by $\bF$, $\sG$, and $\bd$ through the second of \eqref{eq:37uu}. Thus, using the definitions of the strain measures, \eqref{eq:56y}, and the second of \eqref{eq:37uu}, we can write the system of partial differential equations \eqref{eq:54o} in terms of the deformation $\bfy$ and the director field $\bd$.

It is worth noting that the tangential vector $\sm$ appearing in the decomposition of $\bM$ in \eqref{eq:5b} does \color{black}not appear in the power \color{black}expenditure or in the equilibrium equations.

\section{Unshearable shells}
For an unshearable shell, the shear vector defined in \eqref{eq:16} vanishes along every evolution process:
\begin{equation}\label{eq:18}
\ss=\bF^\top\bd=\mathbf 0,
\end{equation}
that is, the director $\bd$ is orthogonal to the range of the deformation gradient $\bF=\nabla\bfy$. At a given point $\bfx$, there are only two unit vectors with this property, namely, $\overline\bn(\bfx)$ and $-\overline\bn(\bfx)$. Moreover, $\bd(\bfx)=\bn(\bfx)$ when the shell is in its reference configuration. Thus, if we maintain that the current configuration is connected with the reference configuration by a smooth family of maps satisfying \eqref{eq:18}, we must have
\begin{equation}\label{eq:36e}
\bd=\overline\bn(\bF)=\frac{\bF^\star\bn}{|\bF^\star\bn|},
\end{equation}
a formula which shows that the current configuration of the shell is determined solely by the deformation $\bfy:\mathcal S\to\overline{\mathcal S}$. As a result, the bending tensor $\sK$ defined in \eqref{eq:16} can be expressed in terms of 
$\bF$ and $\nabla\bF$. Indeed, by taking the gradient of $\bF^\top\overline \bn=\mathbf 0$, we obtain $\bF^\top\nabla\overline\bn+\overline\bn\nabla\bF=\mathbf 0$, which implies \color{black}that\color{black}
 \begin{equation}
 	\bF^\top\bG=-\overline\bn(\bF)\nabla\bF,
 \end{equation}
 whence
 \begin{equation}
 \bG=-\bF^{-\top}(\overline\bn(\bF)\nabla\bF).	
 \end{equation}
 Note that, by construction, $\bF^\top\bG$ is a superficial tensor field. In addition, the third-order tensor \color{black}field \color{black}$\nabla\bF=\nabla\nabla\bfy$, regarded as a vector-valued bilinear form, is symmetric on tangential fields (see \cite{silhavyDirectApproachNonlinear2013}), \emph{i.e.}, $\nabla\bF(\sa\otimes\ssb)=\nabla\bF(\ssb\otimes\sa)$ for every pair $\sa$ and $\ssb$ of tangential vectors. Accordingly, for any such pair we have $(\bF^\top\bG)\cdot(\sa\otimes\ssb)=-(\overline\bn\nabla\bF)\cdot(\sa\otimes\ssb)=\overline\bn\cdot(\nabla\bF(\sa\otimes\ssb))=\overline\bn\cdot(\nabla\bF(\ssb\otimes\sa))=(\bF^\top\bG)\cdot(\ssb\otimes\sa)$. Thus\color{black}, \color{black}$\bF^\top\bG$ is symmetric. Likewise, the bending strain $\sK$ introduced in \eqref{eq:16}, \color{black}namely\color{black}
      \begin{equation}
      {
        \sK=\bF^\top\bG-\nabla\bn=-\overline\bn\nabla\bF-\nabla\bn}\color{black},\color{black}
      \end{equation}
      is a symmetric tensor field. Observe also that, granted \eqref{eq:36e}, the projection tensor $\overline\bP_{\bd}$ coincides with the orthogonal projection
 \begin{equation}
 	\overline\bP=\bI-\overline\bn\otimes\overline\bn.
 \end{equation}
For the vector $\sm$ in the first of \eqref{eq:5b} we now have, by \eqref{eq:2} and by \eqref{eq:18}, $
 \sm=(\bM^\top-\sM^\top\bF^\top)\bd=\bM^\top\bd-\sM^\top\bF^\top\bd=\mathbf 0$. Likewise, by \eqref{eq:16hhh}, \eqref{eq:5b}, and \eqref{eq:37uu}, we have
 $
\sg=(\bG^\top-\sG^\top\bF^\top)\bd=\bG^\top\bd-\sG^\top\bF^\top\bd=\mathbf 0$.
Moreover, the scalar $\lambda$ in the decomposition \eqref{eq:5c} vanishes. Accordingly, \eqref{eq:5a}, \eqref{eq:5b}, and \eqref{eq:5c} are now replaced by
\begin{equation}
\bS=\bF\sN+\overline\bn\otimes\sq,\quad \bM=\bF\sM,\quad \bG=\bF\sG,\quad\bl=\bF\sl.	
\end{equation}
Since $\sg=\mathbf 0$, the shear-force vector $\sq$ and the effective shear-force vector $\tilde\sq$ defined in \eqref{eq:21b} coincide\color{black}:\color{black}
\begin{equation}
\tilde\sq=\sq.	
\end{equation}
We find it convenient to introduce the tensor
\begin{equation}
\bN=\bF\sN.	
\end{equation}
The equilibrium equation \eqref{eq:30} can then be written as:
\begin{equation}\label{eq:22}
 {\rm div}\bN+({\rm div}\sq)\overline\bn+\bG\sq+\bb=\mathbf 0.
 \end{equation}
On projecting \eqref{eq:22} on the tangent plane $T_{\bfy(\bfx)}\mathcal S'$ and along $\overline\bn$, and on observing that, since $\bN^\top\overline\bn=\mathbf 0$, $\overline\bn\cdot{\rm div}\bN={\rm div}(\bN^\top\overline\bn)-\nabla\overline\bn\cdot\bN=-\nabla\overline\bn\cdot\bN=-\bG\cdot\bN$, we obtain
 \begin{equation}\label{eq:split}
 \left.
 \begin{aligned}
 	&\overline{\bP}{\rm div}\bN+\bG\sq+\overline\bP\bb=0,\\
 	&{\rm div}\sq-\bG\cdot\bN+\overline\bn\cdot\bb=0.
 	\end{aligned}
 \right.
 \end{equation}
Next, we write the equilibrium equation \eqref{eq:6} as 
 \begin{equation}\label{eq:projection}
 \overline{\bP}{\rm div}\bM+\mathbf c-\bF\sq=\mathbf 0.
 \end{equation}
We multiply both sides of \eqref{eq:projection} by the pseudoinverse $\bF^{-1}$ of $\bF$; on recalling that $\bF^{-1}\bF=\sP$, that $\bF\bF^{-1}=\overline\bP$, and on observing that $\bF^{-1}\overline\bP=\bF^{-1}(\bF\bF^{-1})=\bF^{-1}$, we get
 \begin{equation}\label{eq:46}
 \sq=\bF^{-1}({\rm div}\bM+\mathbf c).	
 \end{equation}
Using \eqref{eq:46}, the reactive shear force $\sq$ appearing in \eqref{eq:split} can be eliminated, so that the following system is arrived at:
\begin{equation}\label{eq:41}
{
\left.\begin{aligned}	
	&\overline\bP{\rm div}\bN+\bG\bF^{-1}({\rm div}\bM+\mathbf c)+\overline\bP\bb=\mathbf 0,\\
	&{\rm div}(\bF^{-1}({\rm div}\bM+\mathbf c))-\bG\cdot\bN+\overline\bn\cdot\bb=0.
\end{aligned}\right.
}
\end{equation}
The above system is supplemented by constitutive equations for $\bN$ and $\bM$. Bacause of the internal constraint \eqref{eq:18}, the shear vector does not enter the free energy \color{black}density\color{black}, and the constitutive equations \eqref{eq:23} are replaced by
\begin{equation}
	\begin{aligned}
		&\tilde\sN=\partial_{\sE}\tilde\psi(\sE,\sK),\qquad \sM=\partial_{\sK}\tilde\psi(\sE,\sK).
	\end{aligned}
\end{equation}
On recalling that $\sN=\tilde\sN+\sG\sM^\top$, we can write $\bN=\bF\sN=\bF(\partial_{\sE}(\tilde\psi(\sE,\sK)+\sG(\partial_{\sK}\tilde\psi)^\top(\sE,\sK))=\bF\partial_{\sE}\tilde\psi(\sE,\sK)+\bG(\partial_{\sK}\tilde\psi)^\top(\sE,\sK)$. Thus, we obtain the constitutive equations
\begin{equation}\label{eq:52f}
{
\begin{aligned}
&\bN=\bF\partial_{\sE}\tilde\psi(\sE,\sK)+\bG(\partial_{\sK}\tilde\psi)^\top(\sE,\sK),\\
&\bM=\bF\partial_{\sK}\tilde\psi(\sE,\sK).
\end{aligned}
}	
\end{equation}
For a transversely-isotropic shell, a standard choice for the free energy is
\begin{equation}
\tilde\psi(\sE,\sK)=	\frac 1 2 h\mathbb C(	\sE-\sE_0)\cdot(\sE-\sE_0)+\frac 12 \frac {h^3}{12}\mathbb C(\sK-\sK_0)\cdot(\sK-\sK_0),
\end{equation}
where $\boldsymbol{\mathsf E}_0$ and $\boldsymbol{\mathsf K}_0$ are given pre-strains, and where the fourth-order tensor $\mathbb C$ maps superficial tensors into superficial tensors according to the following rule:
\begin{equation}
\mathbb C\sA=2\mu\sA	+\frac{2\mu\lambda}{2\mu+\lambda}({\rm tr}\sA)\sP,
\end{equation}
where $\mu$ and $\lambda$ are the Lam\'e moduli, related to the Young's modulus and Poisson coefficient by the relations
\begin{equation}\label{eq:60h}
\mu=\frac E {2(1+\nu)},\qquad \lambda=\frac{\nu E}{(1+\nu)(1-2\nu)}.	
\end{equation}
Then, 
the constitutive equations \eqref{eq:52f} take the form\urldef\myurl\url{https://uniroma3-my.sharepoint.com/personal/gitomassetti_os_uniroma3_it/_layouts/OneNote.aspx?id=%2Fpersonal%2Fgitomassetti_os_uniroma3_it%2FDocuments%2FNotebooks%2FJournal&wd=target%281%20Year%202022%2FDec%202022.one%7CA7482B2A-331E-4705-9160-6E59866C2EAC%2FConstitutive%20equations%7C946DE726-B3B1-4D4D-96A2-B30123356D10%2F%29}
\mynote{
\myurl}
\begin{equation}\label{eq:55f}
\begin{aligned}
	&\bN=\bF\mathring\sN+\bG\mathring\sM+h\bF\mathbb C\sE+\frac {h^3}{12}\bG\mathbb C\sK,\\
	&\bM=\bF\mathring\sM+\frac{h^3}{12}\bF\mathbb C\sK\color{black},\color{black}
\end{aligned}	
\end{equation}
\color{black}with\color{black}
\begin{equation}
\begin{aligned}
&\mathring\sN=-h\mathbb C\sE_0-\nabla\bn\frac{h^3}{12}\mathbb C\sK_0,\\
&\mathring\sM=-\frac {h^3}{12}\mathbb C\sK_0.
\end{aligned}
\end{equation}

\section{Linearization}
\subsection{Setup} We consider an $\varepsilon$-parametrized family of deformations
\begin{equation}\label{eq:51f}
	\bfy(\bfx)=\bfx+\varepsilon\mathone\bu(\bfx),
\end{equation}
and for convenience we introduce a splitting 
\begin{equation}\label{eq:76bbb}
\varepsilon\mathone\bu=\bu=w\bn+\sv,\qquad \sv\cdot\bn=0.
\end{equation}
of the displacement $\bu=\varepsilon\mathone\bu$ into its normal component $w$ and tangential component $\sv$. Next, we assume that
\begin{equation}\label{eq:57g}
	\sup_{\bfx\in\mathcal S}\Big(\frac{|\bu(\bfx)|}{{\rm diam}(\mathcal S)}+|\nabla\bu(\bfx)|\Big)\color{black}\ll \color{black}1,
\end{equation}
that is, the displacement $|\bu|$ is small compared to the characteristic length of the reference domain $\mathcal S$ and $|\nabla\bu|$ is small as well. We assume that the body couple vanishes and that the body force admits the following expansion:
\begin{equation}
\bb=\mathring\bb+\varepsilon\stackrel 1 {\bb}+o(\varepsilon).
\end{equation}
Likewise, we assume that the tensors $\bN$ and $\bM$ admit the following expansions:
\begin{equation}\label{eq:61f}
\bN=\mathring\sN+\varepsilon\mathone\bN+o(\varepsilon),\qquad \bM=\varepsilon\mathone\bM+o(\varepsilon)\color{black},\color{black}	
\end{equation}
with $\mathring\sN$ and $\mathring\bb$ satisfying the system
\begin{equation}\label{eq:56f}
\left.\begin{aligned}	
	\sP({\rm div}\mathring\sN+\mathring\bb)&=\mathbf 0,\\
	-\nabla\bn\cdot\mathring\sN+\bn\cdot\mathring\bb&=0,
\end{aligned}\right.
\end{equation}
which follows by enforcing the equilibrium equations \eqref{eq:41} for $\varepsilon=0$.

\subsection{The linear problem} By \color{black}linearizing \color{black}the equilibrium equations \eqref{eq:41} and of the constitutive equations \eqref{eq:55f}, we \color{black}aim \color{black}to show that the quantities
\begin{equation}\label{eq:68g}
\mathbf N=\varepsilon\mathone\bN,\qquad \mathbf M=\varepsilon\mathone\bM,\qquad \mathbf b=\varepsilon\mathone\bb,	
\end{equation}
\color{black}obey\color{black} 
\begin{equation}\label{eq:59f}
{\begin{aligned}[b]
	&\sP{\rm div}\mathbf N+\nabla\bn({\rm div}\mathbf M)+\sP\mathbf b=\mathbf 0,\\
	&{\rm div}(\sP{\rm div}\mathbf M)-\nabla\bn\cdot\mathbf N +\mathring\sN\cdot(\nabla\nabla w-\nabla\bn\nabla\sv-\sv\nabla\nabla\bn)\\
	&\qquad +\mathring\bb\cdot(\nabla\bn\sv-\nabla w)+\bn\cdot\mathbf b=0,
\end{aligned}}	
\end{equation}
\color{black}with\color{black}
\begin{align}
&\!\!\mathbf N=\nabla\bu\mathring\sN+h\Big(2\mu\boldsymbol\varepsilon+\frac{2\mu\lambda}{2\mu+\lambda}({\rm tr}\boldsymbol\varepsilon)\sP\Big)+\frac{h^3}{12}\nabla\bn\Big(2\mu\boldsymbol\kappa+\frac{2\mu\lambda}{2\mu+\lambda}({\rm tr}\boldsymbol\kappa)\sP\Big),\nonumber\\
&\!\!\mathbf M=\frac{h^3}{12}\Big(2\mu\boldsymbol\kappa+\frac{2\mu\lambda}{2\mu+\lambda}({\rm tr}\boldsymbol\kappa)\sP\Big),\label{eq:62f}
\end{align}
where
\begin{equation}\label{eq:63f}
{
\begin{aligned}
&\boldsymbol\varepsilon=\frac 12(\sP\nabla\sv+\nabla\sv^\top\sP)+w\nabla\bn,\\
&\boldsymbol\kappa=-\sP\nabla\nabla w+\nabla\bn\nabla\sv+\nabla\sv^\top\nabla\bn+w(\nabla\bn)^2+\sP\,(\sv\nabla\nabla\bn),\OK
	\end{aligned}
	}
\end{equation}
are the linear strains.

 In the special case when $\mathring\sN=\mathbf 0$, the equilibrium equations \eqref{eq:59f} take the form:
\begin{equation}\label{eq:59g}
\begin{aligned}
	&\sP{\rm div}\mathbf N+\nabla\bn({\rm div}\mathbf M)+\sP\mathbf b=\mathbf 0,\\
	&{\rm div}(\sP{\rm div}\mathbf M)-\nabla\bn\cdot\mathbf N+\bn\cdot\mathbf b=0.
\end{aligned}	
\end{equation}
\subsection{Derivation of \eqref{eq:59f}--\eqref{eq:63f}} Observing that $\nabla\bfx=\sP$, from \eqref{eq:51f} we obtain
\begin{equation}\label{eq:56e}
\bF=\sP+\nabla\bu=\sP+\varepsilon\nabla\mathone\bu.
\end{equation}
For the normal vector $\overline\bn$ we assume the expansion
\begin{equation}\label{eq:74}
\overline\bn(\varepsilon)=\bn+\varepsilon\bphi+o(\varepsilon).
\end{equation}
Since $|\overline\bn|=1$, the vector $\bphi$ must be perpendicular to $\bn$:
\begin{equation}
	\bphi\cdot\bn=0.
\end{equation}
Thus, $\sP\bphi=\bphi$ and hence the requirement $\bF^\top\overline\bn=\mathbf 0$ yields
\begin{equation}\label{eq:60e}
\bphi=-\nabla\mathone\bu{}^\top\bn.
\end{equation}
It then follows that the projection $\overline\bP=\bI-\overline\bn\otimes\overline\bn$ admits the expansion $\overline{\bP}(\varepsilon)=\sP-\varepsilon(\bphi\otimes\bn+\bn\otimes\bphi)+o(\varepsilon)$, which, in view of \eqref{eq:60e}, yields
\begin{equation}\label{eq:76f}
\overline{\bP}(\varepsilon)\doteq\sP+\varepsilon(\nabla\mathone\bu^\top\bn\otimes\bn+\bn\otimes\bn\nabla\mathone\bu),
\end{equation}
where the symbol $\doteq$ denotes equality up to the order $\varepsilon$. Furthermore, for the tensor $\bG=\nabla\overline\bn$ we have the expansion $\bG\doteq\nabla\bn+\varepsilon\nabla\bphi$, which entails, because of \eqref{eq:60e},
\begin{equation}\label{eq:77f}
\begin{aligned}
\bG\doteq\nabla\bn-\varepsilon\nabla(\nabla\mathone\bu{}^\top\bn).
\end{aligned}
\end{equation}
\color{black}Using \color{black}\eqref{eq:56e} and \eqref{eq:77f}, \color{black}and noting \color{black}that $\sP\nabla\bn=\nabla\bn$, we obtain the expansions 
\begin{equation}\label{strains}
\begin{aligned}
	\sE&=\frac 1 2(\bF^\top\bF-\sP)=\frac 1 2((\sP+\varepsilon\nabla\mathone\bu{}^\top)(\sP+\varepsilon\nabla\mathone\bu)-\sP)\doteq\varepsilon\mathone\sE,\\
	\sK&=\bF^\top\bG-\nabla\bn\doteq(\sP+\varepsilon\nabla\mathone\bu)^\top\color{black}(\nabla\bn- \varepsilon\nabla(\nabla\mathone\bu{}^\top\bn))-\nabla\bn\doteq\varepsilon\mathone\sK,
\end{aligned}
\end{equation}
for the strain tensors $\sE$ and $\sK$ defined in \eqref{eq:16},
\color{black}with\color{black}
\begin{equation}\label{strains2}
\begin{aligned}
&\mathone\sE=\frac 1 2(\sP\nabla\mathone\bu+\nabla\mathone\bu{}^\top\sP)\quad\text{and}\quad\mathone\sK=-\sP\nabla(\nabla\mathone\bu{}^\top\bn)+\nabla\mathone\bu{}^\top\nabla\bn.
	\end{aligned}
\end{equation}
We observe that both $\mathone\sE$ and $\mathone\sK$ vanish on a rigid displacement. Indeed, any such displacement has the representation
\begin{equation}
	\mathone\bu_{\rm R}(\bfx)=\mathone \bu_0+\mathone{\boldsymbol W}\bfx,
\end{equation}
with $\mathone{\boldsymbol W}=-\mathone{\boldsymbol W}{}^\top$, so that 
\begin{equation}\label{gradrigid}
\nabla\mathone\bu_{\rm R}=\mathone{\boldsymbol W}\nabla\bfx=\mathone{\boldsymbol W}\sP.	
\end{equation}
Substitution of \eqref{gradrigid} into \eqref{strains2} yields $\mathone\sE=\frac 1 2(\sP\mathone{\boldsymbol{W}}\sP+\sP\mathone{\boldsymbol{W}}{}^\top\sP)	=\mathbf 0$ and $\mathone\sK=-\sP\mathone{\boldsymbol W}{}^\top\nabla\bn+\sP\mathone{\boldsymbol W}{}^{\top}\nabla\bn=\mathbf 0$, as claimed.

On using the decomposition \eqref{eq:76bbb}, we can \color{black}write\color{black}
\begin{equation}
\varepsilon\mathone\sE=\frac 12 (\sP\nabla\sv+\nabla\sv{}^\top\sP)+w\nabla\bn=:\boldsymbol\varepsilon.
\end{equation}
Moreover, noting that $\varepsilon\nabla\mathone\bu{}^\top\bn=\varepsilon\nabla(\mathone\bu\cdot\bn)-\varepsilon\nabla\bn\mathone\bu=\varepsilon\nabla(\mathone\bu\cdot\bn)-\varepsilon\nabla\bn\sP\mathone\bu=\nabla w-\nabla\bn\sv$, so that $-\varepsilon\sP\nabla(\nabla\mathone\bu{}^\top\bn)=-\sP\nabla\nabla w+\nabla\bn\nabla\sv+\sP(\sv\nabla\nabla\bn)$, and noting also that $\varepsilon\nabla\mathone\bu{}^\top\nabla\bn=w(\nabla\bn)^2+(\nabla w\otimes\bn)\nabla\bn+\nabla\sv^\top\nabla\bn=w\nabla\bn^2+\nabla\sv^\top\nabla\bn$, we obtain
\begin{equation}
\varepsilon\mathone\sK=-\sP\nabla\nabla w+\nabla\bn\nabla\sv+\nabla\bn\nabla\sv+w\nabla\bn^2+\sP(\sv\nabla\nabla\bn)=:\boldsymbol{\kappa}.
\end{equation}
On recalling \eqref{eq:61f}, the linearization of the constitutive equations \eqref{eq:55f} yields
\begin{equation}
\bN\doteq\mathring\sN+\varepsilon\mathone\bN,\qquad \bM\doteq\varepsilon\mathone\bM,	
\end{equation}
\color{black}with\color{black}
\begin{equation}
\begin{aligned}
&\mathone\bN=\nabla\mathone\bu\mathring\sN+h\mathbb C\mathone\sE+\frac {h^3}{12}\nabla\bn\mathbb C\mathone\sK,\\
&\mathone\bM=\frac{h^3}{12}\mathbb C\mathone\sK.
\end{aligned}
\end{equation}
We can now substitute the above expansions into the equilibrium equations \eqref{eq:41}. At order zero we obtain \eqref{eq:56f}. Taking into account the equilibrium equations at order zero, we obtain, at \color{black}first \color{black}order:
\begin{equation}\label{eq:58f}
\begin{aligned}
	&\sP{\rm div}\mathone\bN+\nabla\bn{\rm div}\mathone\bM+\sP\mathone\bb=\mathbf 0,\\
	&\sP{\rm div}{\rm div}\mathone\bM-\nabla\bn\cdot\mathone\bN+\bn\cdot\mathone\bb+\mathring\sN\cdot\nabla(\nabla\mathone\bu{}^\top\bn)-\nabla\mathone\bu{}^\top\bn\cdot\mathring\bb=\mathbf 0. 
\end{aligned}	
\end{equation}
Multiplying the first of \eqref{eq:58f} by $\varepsilon$ and using \eqref{eq:68g} we obtain the first of \eqref{eq:59f}. Multiplying the second of \eqref{eq:58f} by $\varepsilon$ and observing that $\nabla{\mathone{\boldsymbol{u}}}{}^\top\boldsymbol{n}=\nabla w-\nabla\boldsymbol{n}^\top\mathone\bu=\nabla w-\nabla\boldsymbol{n}\mathone\bu=\nabla w-\nabla\bn\sv$ we obtain the second of \eqref{eq:59f}.

\subsection{A more general case}
It is also possible to derive linearized equations with a non-vanishing initial bending moment and body couple. First, the linearization of the constitutive equations \eqref{eq:55f} yields
\begin{equation}
	\bN\doteq\mathring\sN+\varepsilon \mathone\bN,\qquad \bM\doteq\mathring\sM+\varepsilon\mathone\bM,
\end{equation}
where
\begin{equation}
\begin{aligned}
&\mathone\bN=\nabla\mathone\bu\mathring\sN-(\nabla(\nabla\mathone\bu{}^\top\bn)+\nabla\mathone\bu\nabla\bn)\mathring\sM+h\mathbb C\mathone\sE+\frac {h^3}{12}\nabla\bn\mathbb C\mathone\sK,\\
&\mathone\bM=\nabla\mathone\bu\mathring\sM+\frac{h^3}{12}\mathbb C\mathone\sK.
\end{aligned}
\end{equation}
The equilibrium equations at \color{black}zeroth order \color{black}are
\begin{equation}\label{eq:56ee}
\left.\begin{aligned}	
	&\sP{\rm div}\mathring\sN+\nabla\bn({\rm div}\mathring\sM+\mathring\ssc)+\sP\mathring\bb=\mathbf 0,\\
	&{\rm div}(\sP{\rm div}\mathring\sM+\mathring{\ssc}))-\nabla\bn\cdot\mathring\sN+\bn\cdot\mathring\bb=0.
\end{aligned}\right.
\end{equation}
Next, we need to compute the linearization of the pseudo-inverse. At first order, we have the expansion
\begin{equation}\label{eq:75b}
\begin{aligned}
\bF^{-1}(\varepsilon)&=\sP+\varepsilon\bL+o(\varepsilon),\\
\end{aligned}
\end{equation}
for some tensor $\bL$. To identify $\bL$, we recall that $\bF^{-1}\bF=\sP$. Then, on combining \eqref{eq:56e} and \eqref{eq:75b} we obtain $\bF^{-1}(\varepsilon)\bF(\varepsilon)=\sP+\varepsilon(\bL\sP+\sP\nabla\mathone\bu)+o(\varepsilon)$. Thus,  
\begin{equation}\label{eq:77}
\bL\sP=-\sP\nabla\mathone\bu.	
\end{equation}
Moreover, recalling that the pseudo-inverse $\bF^{-1}$ annihilates $\bd=\overline\bn$, we write $\mathbf 0=\bF^{-1}(\varepsilon)\overline\bn(\varepsilon)=(\sP+\varepsilon\bL+o(\varepsilon))(\bn+\varepsilon\bphi+o(\varepsilon))=\varepsilon(\bL\bn-\nabla\mathone\bu{}^\top\bn)+o(\varepsilon)$, to obtain
\begin{equation}\label{eq:78}
\bL\bn=\nabla\mathone\bu{}^\top\bn.	
\end{equation}
Since $\bL=\bL(\sP+\bn\otimes\bn)=\bL+(\bL\bn)\otimes\bn$, on combining \eqref{eq:77} and \eqref{eq:78} we obtain the following expression for the first-order term in the expansion of $\bF^{-1}(\varepsilon)$:
\begin{equation}\label{eq:75d }
\bL=-\sP\nabla\mathone\bu+\nabla\mathone\bu{}^\top\bn\otimes\bn.	
\end{equation}
Substituting in to the equilibrium equations we obtain, at \color{black}first \color{black}order,
\begin{align}
	&\sP{\rm div}\mathone\bN+\nabla\bn({\rm div}\mathone\bM+\mathone\bc)+\sP\mathone\bb\nonumber\\
	&\qquad+(-\nabla(\nabla\mathone\bu{}^\top\bn)+\nabla\mathone\bu{}^\top\bn\otimes\bn-\sP\nabla\mathone\bu)({
	\rm div}\mathring\sM+\mathring\ssc)=\mathbf 0.\nonumber\\
	&{\rm div}(\sP({\rm div}\mathone\bM+\mathone\bc))-\nabla\bn\cdot\mathone\bN+\bn\cdot\mathone\bb+{\rm div}((\nabla\mathone\bu{}^\top\bn\otimes\bn-\sP\nabla\mathone\bu)({\rm div}\mathring\sM+\mathring\ssc))\nonumber\\
	&\qquad+\mathring\sN\cdot\nabla(\nabla\mathone\bu{}^\top\bn)-(\bn\otimes\mathring\bb)\cdot\nabla\mathone\bu=\mathbf 0. \label{eq:58e}
\end{align}

\section{Application: free vibrations of a pressurized spherical shell}

As an application of the theory, we deduce and solve the equations that govern the free vibrations of a pressurized spherical shell. For a generic geometry, incremental equations governing equilibrium for small departures from a pre-stressed reference configuration may be found in \cite[p.\ 452]{timoshenko1961theory} and in \cite{zubovTheorySmallDeformations1976,zubovCompatibilityEquationsStress1985}. A set of equations obtained by a systematic linearization of non-linear theory has been carried out more recently in \cite{altenbachVibrationAnalysisNonlinear2014}. For a spherical shell, incremental equilibrium equations can also be found in \cite[p. 262]{flueggeStatikUndDynamik1962}. More recently, the equations that govern axisymmetric oscillations of pressurized spherical shell have been derived and studied by \cite{kuoSmallOscillationsPressurized2015}, where the contribution of the pre-stress induced by the internal pressure is accounted for by \color{black}an \emph{ad hoc} argument\color{black}.

To maintain a manageable complexity in the equations, coordinate-based treatments are \color{black}usually restricted \color{black} to axisymmetric solutions, and often neglect bending contributions \color{black}(see for instance \cite{bakerAxisymmetricModesVibration1961a,flueggeStatikUndDynamik1962,randVibrationsFluidFilled1967})\color{black}. On the contrary, our concise notation alleviates the need for such assumptions. Still, to underscore the key aspects of the linearization process, we will initially set aside bending moments, introducing them subsequently.\color{black} 

\subsection{The reference state} We suppose that the body admits an equilibrium state, which we refer to as \color{black}the \color{black}\emph{reference state}, where its material elements occupy a spherical surface of radius $R$, under the action of the body-force \color{black}field\color{black}
\begin{equation}\label{eq:92f}
\mathring \bb=p\bn,\OK
\end{equation}
and a null body-couple field. We assume that in the reference state 
\begin{equation}\label{eq:61e}
	\mathring{\sN}=\frac{pR}2\sP,\qquad\mathring{\sM}=\mathbf 0.\OK
      \end{equation}
To verify that the body-force field \eqref{eq:92f} equilibrates the stress fields \eqref{eq:61e} we observe that, since\color{black}, \color{black}for a sphere of radius $R$\color{black},\color{black}
\begin{equation}\label{eq:91f}
\nabla\bn=\frac 1 R\sP,\OK
\end{equation}
we have ${\rm div}\sP={\rm div}(\bI-\bn\otimes\bn)=-(\nabla\bn)\bn]-(\sP\cdot\nabla\bn)\bn=-(2/R)\bn$, and\color{black}, \color{black}hence\color{black},\color{black}
\begin{equation}\label{eq:64f}
{\rm div}\mathring\sN=\frac{pR}2{\rm div}{\sP}=-p\bn=-\mathring\bb,	\OK
\end{equation}
and
\begin{equation}
  \mathring\sN\cdot\sP=p.\OK
  \end{equation}
Thus the equilibrium equations \eqref{eq:56f} are satisfied.

\subsection{Perturbed states}
We suppose that the body force and body couple in a generic \color{black}state are given \color{black}by:
\begin{equation}\label{eq:5}
\bb=p\,\bF^\star\bn-h\rho\bu'',\qquad\bc=\mathbf 0.\OK
\end{equation}
We recall that $\bF^\star$ is the cofactor of $\bF$, and that $\bF^\star(\bfx)\bn(\bfx)$ is the vector pointing in the direction of $\overline\bn(\bfx)$, \color{black}with \color{black}intensity \color{black}being \color{black}the ratio between an infinitesimal surface element placed at $\bfx$ in the reference configuration and the area of its image under the deformation $\bfy$. 

We \color{black}next \color{black}evaluate \eqref{eq:5} for small departures from the reference state. To begin with, we observe that
\begin{equation}\label{eq:113ggg}
	\bF^\star\bn={\rm det}(\bF+\overline\bn\otimes\bn)\overline\bn.\OK
\end{equation}
One way to verify \eqref{eq:113ggg} is as follows. Let $\sa$ and $\ssb$ two tangent vectors in the reference configuration such that $\bn=\sa\times\ssb$. 
Then, $\bF\sa$ and $\bF\ssb$ are tangent to $\overline{\mathcal S}$ and are linearly independent, since $\bF$ is injective. Thus, $\bF^\star\bn=\bF^\star(\sa\times\ssb)=\bF\sa\times\bF\ssb$ is parallel to $\overline\bn$. \color{black}Also, \color{black}for $\widetilde\bF=\bF+\overline{\bn}\otimes\bn$, we have $\widetilde\bF\bn=\overline\bn$, $\widetilde\bF\sa=\bF\sa$, and $\widetilde\bF\ssb=\bF\ssb$. Since $\bF^\star\bn$ is parallel to $\overline\bn$, we have $ \bF^\star\bn=(\overline\bn\cdot\bF^\star\bn)\overline\bn$. 
Moreover, $\overline\bn\cdot\bF^\star\bn=\overline\bn\cdot\bF\sa\times\bF\ssb=\widetilde\bF\bn\cdot\widetilde\bF\sa\times\widetilde\bF\ssb=\det(\widetilde\bF)(\bn\cdot\sa\times\ssb)=\det(\bF+\overline\bn\otimes\bn)$, whence \eqref{eq:113ggg}.

Now, \color{black}we can use \color{black}\eqref{eq:56e} and \eqref{eq:74} \color{black}to \color{black}write 
${\rm det}(\bF+\overline\bn\otimes\bn)$ 
$=\det(\sP+\bn\otimes\bn+\varepsilon\nabla\mathone\bu+(\overline\bn-\bn)\otimes\bn)$
$=\det(\bI+\varepsilon\nabla\mathone\bu+(\overline\bn-\bn)\otimes\bn)$ 
$=\det(\bI+\varepsilon(\nabla\mathone\bu+\bphi\otimes\bn)+o(\varepsilon))$
$\doteq 1+\varepsilon \rm tr(\nabla\mathone\bu)$ 
$=1+\varepsilon\sP\cdot\nabla\mathone\bu$ 
$=1+\varepsilon\sP\cdot(\nabla\mathone\sv+\mathone w\nabla\bn+\bn\otimes\nabla \mathone w)$ 
$=1+\varepsilon({\rm div}\mathone\sv+2\mathone w/R)$\color{black}; thus,\color{black}
\begin{equation}\label{eq:114bbb}
\det(\bF+\overline\bn\otimes\bn)\doteq 1+\varepsilon\Big({\rm div}\mathone\sv+2\frac {\mathone w}	 R\Big).\OK
\end{equation}
Moreover, recalling again \eqref{eq:74}--\eqref{eq:60e} and using \eqref{eq:91f}, we \color{black}find that \color{black}
$\overline\bn\doteq\bn+\varepsilon(-\nabla\mathone\bu{}^\top\bn)$ 
$=\bn+\varepsilon(-\nabla\mathone w+\nabla\bn\mathone\sv)$, whence
\begin{equation}\label{eq:115bbb}
\overline\bn\doteq\bn+\varepsilon\Big(-\nabla\mathone w+\frac 1 R\mathone\sv\Big).\OK
\end{equation}
By combining \eqref{eq:5}--\eqref{eq:115bbb} we obtain $\bb\doteq\mathring\bb+\varepsilon\mathone\bb$ where $\mathring\bb$ is given by \eqref{eq:92f} and $\mathone\bb=p\big({\rm div}\mathone\sv+2{\mathone w}/R\big)\bn+p\big(-\nabla\mathone w+\mathone\sv/R\big)-h\varrho\mathone\bu{}''$. Accordingly, the force increment in \eqref{eq:59g} \color{black}is\color{black}
\begin{equation}\label{eq:95f}
\mathbf b=\varepsilon\mathone\bb=p\Big({\rm div}\sv+2\frac {w} R\Big)\bn+p\Big(-\nabla w+\frac 1 R\sv\Big)-h\varrho w''\bn-h\varrho\sv''.	\OK
\end{equation}

\subsection{Incremental equilibrium equations\color{black}: \color{black}the case of null bending moment}
We now write the incremental equilibrium equations \eqref{eq:59f} specialized to the case when the shell does not resist bending. We have from \eqref{eq:95f} that the tangential component of the force increment $\mathbf b$ is
\begin{equation}\label{eq:103f}
\sP\mathbf b=-\varrho h\sv''-p\nabla w+\frac p R\sv.\OK
\end{equation}
Furthermore, recalling that $\mathring\bb$ is parallel to $\bn$, and observing that both $\nabla\bn \sv$ and $\nabla w$ are orthogonal to $\bn$, we have from \eqref{eq:95f},
\begin{equation}\label{eq:106g}
\bn\cdot\mathbf b+ \mathring\bb\cdot(\nabla\bn\sv-\nabla w)\color{black}=p\operatorname{div}\sv+2p\frac w R -h\varrho\sv''.\OK
\end{equation}
Moreover, by \eqref{eq:61e},
\begin{align}
 \mathring\sN\cdot(\nabla\nabla w-\nabla\bn\nabla\sv-\sv.\nabla\nabla\bn)&{}=\frac{pR}2\sP\cdot\Big(\nabla\nabla w-\frac 1 R\sP \nabla\sv-\frac 1 R\sv\nabla\sP\Big)\nonumber\\
&{}=\frac{p}2R\Delta w-\frac p 2{\rm div}\sv\color{black}.\label{eq:107g}\color{black}
\end{align}
Here we have used the \color{black}relation \color{black}$\sP\cdot(\sv\nabla\sP)=0$. Indeed, given a tangent vector $\sa$, we have $(\sv\nabla\sP)\sa=-(\sv\cdot\nabla\bn\sa) \bn=(\bn\otimes\nabla\bn\sv)\sa$. Thus, $\sv\nabla\sP=\bn\otimes\nabla\bn\sv$. Therefore, $\sP\cdot(\sv\nabla\sP)=\sP\cdot\bn\otimes\nabla\bn\sv=\sP\bn\cdot\nabla\bn^\top\sv=0$. By \eqref{eq:103f}, \eqref{eq:106g}\color{black}, \color{black}and \eqref{eq:107g}, the equilibrium equations \eqref{eq:59f} with $\mathbf M=\mathbf 0$ take the form:
\begin{equation}\label{eq:66e}
\begin{aligned}
	&\sP{\rm div}\mathbf N-p\nabla w+\frac p R\sv\color{black}=h\varrho\sv'',\OK\\
	&-\nabla\bn\cdot\mathbf N+\frac {p}2R\Delta w+\frac p 2 {\rm div}\sv+2p\frac w R=h\varrho w''.\OK
\end{aligned}	
\end{equation}
We next employ the constitutive equations \eqref{eq:62f}\color{black}, \color{black}with $\mathbf M=\mathbf 0$\color{black}, \color{black}together with the definitions \eqref{eq:63f} \color{black}yield\color{black}
\begin{align}
\mathbf N
={}&\frac {pR} 2\nabla\sv+\mu h(\sP\nabla\sv+\nabla\sv^\top\sP)+\tilde\lambda h({\rm div}\sv)\sP\nonumber\\
&+\Big(\frac {pR} 2
+2(\mu+\tilde\lambda) h\Big)\frac w R\sP+\frac {pR}2\bn\otimes\nabla w,\label{eq:122bbb}
\end{align}	
where we have \color{black}introduced\color{black}
\begin{equation}
\tilde\lambda=\frac{2\mu\lambda}{2\mu+\lambda}.\OK
\end{equation}
We \color{black}use \color{black}the \color{black}standard \color{black}notation
\begin{equation}
\Delta\sv={\rm div}\nabla\sv
\end{equation}
\color{black}to denote the Laplacian of $\sv$. \color{black}It is shown in Appendix A.6 that the following identities hold:
\begin{equation}\label{eq:125}
\begin{aligned}
	&\sP{\rm div}(\sP\nabla\sv)=\sP\Delta\sv+\frac 1 {R^2}\sv,\OK\\
	&\sP{\rm div}(\nabla\sv^\top\sP)=\nabla{\rm div}\sv+\frac 1 {R^2}\sv,\OK\\
	&\sP{\rm div}(({\rm div}\sv)\sP)=\nabla{\rm div}\sv,\OK\\
	&\sP{\rm div}\Big(\frac w R\sP\Big)=\frac 1 R\nabla w,\OK\\
	&\sP{\rm div}(\bn\otimes\nabla w)=\frac 1 R \nabla w.\OK
	\end{aligned}
\end{equation}
Using \eqref{eq:122bbb} and \eqref{eq:125}, we can write:
\begin{align}
\sP{\rm div}\mathbf N
={}&\Big(\frac {p}2+\mu \frac h R\Big)R\sP\Delta\sv+(\mu+\tilde\lambda)\frac h R (R\nabla{\rm div}\sv)\nonumber\\
&+\Big(p+2(\mu+\tilde\lambda)\frac h R\Big)\nabla w+2\mu \frac h R \frac \sv R.\label{eq:102g}
\end{align}
Furthermore,
\begin{equation}\label{eq:114g}
	-\nabla\bn\cdot\mathbf N=-\frac 1 R\sP\cdot\mathbf N=-\Big(\frac p 2+2(\mu+\tilde\lambda)\frac h R\Big){\rm div}\sv-\Big(p+4(\mu+\tilde\lambda)\frac h R\Big)\frac w R.\OK
\end{equation}
We note that, since $\mathring{\bb}$ is parallel to $\bn$, the last term on the right-hand side of the second of \eqref{eq:59f} vanishes.

By making use of \eqref{eq:122bbb}--\eqref{eq:114g}, the equilibrium equations \eqref{eq:66e} take the form
\urldef\myurl\url{https://uniroma3-my.sharepoint.com/personal/gitomassetti_os_uniroma3_it/_layouts/OneNote.aspx?id=%2Fpersonal%2Fgitomassetti_os_uniroma3_it%2FDocuments%2FNotebooks%2FJournal&wd=target%281%20Year%202022%2FDec%202022.one%7CA7482B2A-331E-4705-9160-6E59866C2EAC%2FUntitled%20Page%7C9D0A692D-6E5A-EB4A-8D8F-65238EDCCF7C%2F%29}
\mynote{\myurl}
\begin{equation}\label{eq:98h}
\begin{aligned}
	&\Big(\frac h R\mu+\frac {p}2\Big)\Big(R\sP\Delta\sv+2\frac\sv R\Big)+\frac h R(\mu+\tilde\lambda)\nabla\Big(R{\rm div}\sv+2 w\Big)=\varrho h\sv'',\OK\\
	&\frac p 2\Big(R\Delta w+2\frac w R\Big)-2\frac h R(\mu+\tilde\lambda)\Big({\rm div}\sv +2\frac w R\Big)=\varrho h w''.\OK
	\end{aligned}
\end{equation}
The structure of \eqref{eq:98h} \color{black}is straightforward\color{black}. The left-hand side of each equation comprises two terms, each one being the product of a coefficient --- dependent on both the properties of the shell and the pressure --- and a term which consists in the action of differential operators on the normal and tangential components of the displacement field, and which vanishes for infinitesimal rigid displacements.

\subsection{Coordinate expression for axisimmetric motions.}
Using a system of spherical coordinates $(\theta,\phi)$ described in Appendix A.5, we represent a time-dependent axisymmetric displacement field $\bu(\theta,\phi,t)$ as the sum
\begin{equation}
\color{black}\bu(\theta,\phi,t)=\sv(\theta,\phi,t)+w(\theta,t)\bn(\theta)
\end{equation}
of a normal component $w(\theta,t)\bn(\theta)$ and a tangential component 
\begin{equation}
\sv(\theta,\phi,t)=v_{\langle\theta\rangle}(\theta,t)\sa_{\langle\theta\rangle}(\theta),
\end{equation}
where $\sa_{\langle\theta\rangle}(\theta)$ is the first element of the physical basis.

Using the results from Appendix A.5 concerning the coordinate representation of the differential operators $\Delta$, $\nabla{\rm div}$, ${\rm div}$, and $\nabla$, and recalling also \eqref{eq:60h}, we find that the first of the equations of motion \eqref{eq:98h} has only one nontrivial component, namely, that in the direction of $\sa_{\langle\theta\rangle}$:
\begin{align}\label{eq:111h}
	&\frac{Eh}{R^2(1-\nu^2)}\Big(\frac{\partial^2 v_{\langle\theta\rangle}}{\partial\theta^2}+\frac{\partial}{\partial\theta}(\cot\theta v_{\langle\theta\rangle})+(1-\nu)v_{\langle\theta\rangle}+(1+\nu)\frac{\partial w}{\partial\theta}\Big)\nonumber\\
	&\qquad +
	\frac{p}{2} \Big(\frac{2}{R} \frac{\partial w}{\partial\theta}+\cot (\theta ) \frac{\partial v_{\langle\theta\rangle}}{\partial\theta}+\frac{\partial^2 v_{\langle\theta\rangle}}{\partial\theta^2}-(\csc (\theta ))^2 v_{\langle\theta\rangle}\Big)
	=\rho h v''_{\langle\theta\rangle}.
\end{align}	
We also find that the second equation takes the form
\begin{equation}\label{eq:112h}
\begin{aligned}
-\frac{Eh}{(1-\nu)R^2}{\Big(v_{\langle\theta\rangle}'+\cot\!\theta\, v_{\langle\theta\rangle}+2 w\Big)}+\frac{p}{2R}\Big(\frac{\partial^2 w}{\partial\theta^2}+\cot\!\theta\,\frac{\partial w}{\partial\theta}+2w\Big)=\rho h w''.
\end{aligned}
\end{equation}
For $p=0$\color{black}, \color{black}\eqref{eq:111h} and \eqref{eq:112h} coincide, respectively, \color{black}with (1) \color{black}and (2) of \cite{kuoSmallOscillationsPressurized2015}. However, the extra terms arising from $p$ are different from those obtained by \cite{kuoSmallOscillationsPressurized2015}. \color{black}In particular, 
in our formulation the initial pre-stress also affects the tangential motion of the shell. However, 
the extra contribution of the normal component of the displacement that appears in our 
equations of motion coincides with that in the corresponding equations of \cite{kuoSmallOscillationsPressurized2015}.\color{black}

\subsection{Solutions using vector spherical harmonics}
We divide both sides of \eqref{eq:98h} by $\mu$ and we introduce the dimensionless parameters $H=h/R$, $P=p/\mu$, $\Lambda=\tilde\lambda/\mu=2\lambda/(2\mu+\lambda)=2\nu/(1-\nu)\in[-1,2]$. Here $\nu\in[-1,1/2]$ is the Poisson ratio. We also introduce the rescaled displacements $\overline w=w/R$, $\overline\sv=\sv/R$, as well as the rescaled coordinates $\overline\bfx=\bfx/R$ and the rescaled time $\color{black}\overline t=t\sqrt{{\rho h R}/{\mu}}$. Dropping overbars, and dividing both sides by $H$ we arrive at the following system:
\begin{equation}\label{eq:99h}
\begin{aligned}
	&\Big(1+\frac 1 2 \frac {P}{H}\Big)(\sP\Delta\sv+2\sv)+(1+\Lambda)\nabla({\rm div}\sv+2 w)={\sv}'',\OK\\
	&\frac 12 \frac P H (\Delta w+2w)-2(1+\Lambda)({\rm div}\sv+2w)=w''.\OK
	\end{aligned}
\end{equation}
We represent the unknowns $w$ and $\sv$ as
\begin{equation}\label{eq:114f}
\begin{aligned}
\sv(\bfx,t)&=\sum_{n\ge 1}\sum_{-n\le m\le n}v^{{\rm b}}_{nm}(t)\ssb_{nm}(\bfx)+v^{\rm c}_{mn}(t)\ssc_{mn}(\bfx)+{\rm c.c.},\\
w(\bfx,t)&=\sum_{n\ge 0}\sum_{-n\le m\le n}	w_{nm}(t)Y_{nm}(\bfx)+{\rm c.c.},
\end{aligned}
\end{equation}
where $Y_{nm}(\bfx)$  are the complex spherical harmonics and $\ssb_{nm}(\bfx)$ and $\ssc_{nm}(\bfx)$ are the complex vector spherical harmonics  vector spherical harmonics on the unit sphere. The symbol c.c. stands for complex conjugate.

Spherical harmonics are a classical tool of \color{black}m\color{black}athematical \color{black}p\color{black}hysics. Their definition may be found, for example, in \cite{morseMethodsTheoreticalPhysics1999}. For the purposes of the present paper, it \color{black}suffices to \color{black}recall that the set $\{Y_{nm},n\ge 0,-n\le m\le m\}$ is an orthonormal (with respect to the $L^2$ scalar product), complete system on the unit sphere, and its elements are eigenfunction\color{black}s \color{black}of the \color{black}Laplace \color{black}operator:
\begin{equation}\label{eq:137bbb}
-\Delta Y_{nm}=s_n^2{Y_{nm}},\qquad s_n=\sqrt{n(n+1)}.
\end{equation}
For $n\ge 1$ and $-n\le m\le n$, the vector spherical harmonics $\ssb_{nm}$, and $\ssc_{nm}$ can be defined on the unit sphere by:
\begin{equation}\label{eq:138bbb}
\ssb_{nm}=\frac 1 {s_n}\nabla Y_{nm},\qquad 	\ssc_{nm}=\bn\times\ssb_{nm}.
\end{equation}
Both $\ssb_{nm}$ and $\ssc_{nm}$ are tangential vector fields. These fields satisfy:
\begin{equation}
\operatorname{div}\ssb_{nm}=-s_{nm}Y_{nm},\qquad \operatorname{div}\ssc_{nm}=0,	
\end{equation}
along with
\begin{equation}\label{eq:140ccc}
-\sP\Delta\ssb_{nm}=s_n^2\ssb_{nm},\qquad -\sP\Delta\ssc_{nm}=s_n^2\ssc_{nm},
\end{equation}
Comprehensive discussions of vector spherical harmonics and their applications may be found in \cite{barreraVectorSphericalHarmonics1985},   \cite{morseMethodsTheoreticalPhysics1999}, and \cite{quartapelleSpectralSolutionThreedimensional1995}. Our notation follows more closely \cite{quartapelleSpectralSolutionThreedimensional1995}. We \color{black}advise \color{black}the reader that  \cite{quartapelleSpectralSolutionThreedimensional1995} define spherical harmonics and vector spherical harmonics 
over the entirety of Euclidean point space $\mathcal E$. We also remark that our definition of $\ssc_{nm}$, although equivalent when the domain of definition is the unit sphere, is different from that given in \cite{quartapelleSpectralSolutionThreedimensional1995}.

Substituting \eqref{eq:114f} into \eqref{eq:99h}, \color{black}invoking \color{black}\eqref{eq:138bbb}--\eqref{eq:140ccc} and using the orthogonality of spherical harmonics and vector spherical harmonics, we find that that $\color{black}w_{00}$ solves
\begin{equation}\label{eq:141hhh}
\frac{{\rm d}^2 w_{00}}{{\rm d} t^2}=\Big(\frac P H-4(1+\Lambda)\Big)w_{00}.
\end{equation}
For $P/H<4(1+\Lambda)$, the solutions of \eqref{eq:141hhh} are oscillatory motions describing isotropic expansion and contraction. For $P/H>4(1+\Lambda)$ solutions increase exponentially in time, a behaviour that is the hallmark of the instability of the reference state. This indicates that internal pressure can indeed foster \color{black}instability\color{black}. Indeed, as the shell expands, the pressure per unit reference area increases, which in turn pushes the shell further outwards.

For $n\ge 1$ and $-n\le m\le n$, the coefficients $w_{mn}(t)$, $v^{\rm b}_{mn}(t)$ solve the following system of differential equations:
\begin{equation}\label{eq:115h}
\begin{aligned}
&\frac{{\rm d}^2 v^{\rm b}_{mn}}{{\rm d} t^2}=\Big(\Big(1+\frac 1 2 \frac P H\Big)(2-s_n^2)-(1+\Lambda)s_n^2\color{black}\Big)v^{\rm b}_{mn}+2(1+\Lambda)s_n\color{black}w_{mn},\\
&\frac{{\rm d}^2 w_{mn}}{{\rm d} t^2}=2(1+\Lambda)s_n v^{\rm b}_{mn}+\Big(\frac P H-\frac 1 2\frac P H s_n^2-4(1+\Lambda)\Big)w_{mn}.
\end{aligned}
\end{equation}
Moreover, $v^{\rm c}_{mn}(t)$ solves: 
\begin{equation}\label{eq:119h}
	\frac{{\rm d}^2 v^{\rm c}_{mn}}{{\rm d} t^2}=\Big(\frac P H\color{black}-\Big(\frac 12\frac P H+1\Big)s_n^2+2\Big)v_{mn}^{\rm c}.
\end{equation}
System \eqref{eq:115h} can be written, in compact form,
\begin{equation}\label{eq:155yyy}
\frac{{\rm d}^2}{{\rm d}t^2}	\mathbbm u_{nm}(t)=\mathbb A_n\mathbbm u_{nm}(t),
\end{equation}
where $\mathbbm u_{mn}=(
v_{mn}^{\rm b},w_{mn})^\top$ and 
\begin{equation}\label{eq:146ppp}
\mathbbm A_n{=}	
\begin{pmatrix}
\displaystyle\Big(1+\frac 1 2 \frac P H\Big)(2-s_n^2)-(1+\Lambda)s_n^2\color{black}
&2(1+\Lambda) s_n 
 & \\[0.5em] 
2(1+\Lambda)s_n&\displaystyle\frac 12\frac P H(2-s_n^2)-4(1+\Lambda)\\
\end{pmatrix}.
\end{equation}
For $n=1$ and $-1\le m\le 1$, the vector spherical harmonics $\ssc_{nm}$ are infinitesimal rotations about an axis going through the origin, and \eqref{eq:119h} reduces to ${{\rm d}^2}v_{nm}^c(t)/{{\rm d}t^2}=0$.   Likewise, for $n=1$, \eqref{eq:146ppp} reduces to
\begin{equation}
\mathbbm A_0=	
\begin{pmatrix}
-2(1+\Lambda)
&2 \sqrt 2 (1+\Lambda)
 & \\ 
2\sqrt{2}(1+\Lambda)&-4(1+\Lambda)\\
\end{pmatrix},
\end{equation}
which is singular matrix, with eigenvalues  and $-6(1+\Lambda)$, $0$. A pair of corresponding eigenvectors is, in the same order, $(1,-\sqrt{2})^\top$ and $(1,1/\sqrt{2})^\top$. The eigenvector corresponding to the null eigenvalue describes a translation with constant velocity. In particular, the null eigenvector corresponds to solutions of the form 
\begin{equation}
\sv(\bfx,t)=	(a_0+b_0 t)\ssb_{1m}(\bfx),\qquad w(\bfx,t)=\frac 1 {\sqrt 2}(a_0+b_0 t)Y_{1m}(\bfx)\bn
\end{equation}
with $m=-1,0,1$. In particular, $m=0$ corresponds to translations parallel to the polar axis, whereas linear combinations of solutions with $m=-1$, and $m=1$ generate translations in the equatorial direction. We next compute:
\begin{align}
\operatorname{tr}(\mathbb A_n)={}&	-(4+n (n+1))\Lambda-2 n (n+1)-2+(2-n (n+1)) P\color{black},\color{black}\nonumber\\
\det(\mathbb A_n)={}&\frac{1}{4} \left(n^2+n-2\right)+16+ (4 + 4 n (1 + n)) P + (-2 + n (1 + n)) P^2\nonumber\\
&+(16 + (8 + 2 n (1 + n)) P)\Lambda.
\end{align}
Note that $\Lambda=2\nu/(1-\nu)$ is an increasing function of $\nu$, ranging from $-1$ (for $\nu=-1$) to $2$ (for $\nu=0.5$). For these extreme values, and for $P\ge 0$, the trace of $\mathbb A_n$ is negative, while the determinant of $\mathbb A_n$ is positive. Since both the trace and the determinant of $\mathbb A_n$ are linear functions of $\Lambda$, we conclude that when $P$ is positive the eigenvalues of $\mathbb A_n$ are negative for $n\ge 2$.

\subsection{Case of non-vanishing bending moment}
It is not difficult to to include bending moments in our analysis. For a spherical geometry, the constitutive equation \eqref{eq:62f} for $\mathbf M$ takes the form:
\begin{align}
\color{black}\mathbf M&=\color{black}\frac {h^3}{12}\Big[2\mu\Big(-\sP\nabla\nabla w+\frac 1 R(\sP\nabla\sv+\nabla\sv^\top\sP)+\frac w {R^2}\sP\Big)\nonumber\\
&\ \ \qquad\qquad \qquad\qquad\qquad\qquad\color{black}\color{black}+\tilde\lambda\Big(-\Delta w+\frac 2 R{\rm div}\sv+2\frac w {R^2}\Big)\sP\Big].
\end{align}
To deduce the equilibrium equations we observe that, compared to the case when $\mathbf M=\mathbf 0$,  the presence of the bending moment affects the first equilibrium equation of \eqref{eq:59f} through the presence of the additional term
\urldef\myurl\url{https://uniroma3-my.sharepoint.com/personal/gitomassetti_os_uniroma3_it/_layouts/OneNote.aspx?id=%2Fpersonal%2Fgitomassetti_os_uniroma3_it%2FDocuments%2FNotebooks%2FJournal&wd=target%280%20Year%202023%2FAug%202023.one%7C02EA3BF8-3F14-0049-8D7F-0673A892FCF5%2FVerification%20of%20eq%3A160bbb%7C9EAAEA33-51CC-8641-A5C0-1654965CCA1A%2F%29}
\mynote{\myurl}
\begin{multline}
	\frac{2}{R} \mathbf{P} \operatorname{div} \mathbf{M} = \frac{1}{6}\left(\frac{h}{R}\right)^3  \left[-(2 \mu+\widetilde{\lambda}) \nabla\left(R^2 \Delta w + 2 \frac{w}{R}\right)\right.\\
	 \left.+ 2(\mu+\widetilde{\lambda}) \nabla(R \operatorname{div} \mathbf{v} + 2 w)\right.
	 \left.+ 2 \mu\left(R \mathbf{P} \Delta \mathbf{v} + 2 \frac{\mathbf{v}}{R}\right)\right].\label{eq:160bbb}
\end{multline}
Likewise, in the second of the equilibrium equations \eqref{eq:59f} the contribution from the bending moment is
\begin{align}
&{\rm div}{\sP{\rm div}\mathbf M}-\frac 1 {R^2}\sP\cdot\mathbf M\nonumber\\
&{}=\quad\!\color{black}\frac{1}{12}\left(\frac{h}{R}\right)^3\left[(2 \mu+\widetilde{\lambda})\left(-R^3 \Delta \Delta w\right)\right.+2 \mu R^2 \operatorname{div}\left(\mathbf{P} \Delta \mathbf{v}\right)\nonumber\\ 
&\qquad\qquad \qquad 
\left.+2(\mu+\widetilde{\lambda}) R^2 \Delta \operatorname{div} \mathbf{v}-4(2 \mu+\widetilde{\lambda}) \frac{w}{R}\right]\nonumber\\
&\ \quad+\frac{1}{6}\left(\frac{h}{R}\right)^3(\mu+2 \widetilde{\lambda})\left(R \Delta w+2 \frac{w}{R}\right)\nonumber\\
&\ \quad-\frac{1}{3}\left(\frac{h}{R}\right)^3 \tilde{\lambda}\left(\operatorname{div} \mathbf{v}+2 \frac{w}{R}\right).
\end{align}

Thus, the equilibrium equations \eqref{eq:98h} are replaced by
\urldef\myurl\url{https://uniroma3-my.sharepoint.com/personal/gitomassetti_os_uniroma3_it/_layouts/OneNote.aspx?id=%2Fpersonal%2Fgitomassetti_os_uniroma3_it%2FDocuments%2FNotebooks%2FJournal&wd=target%280%20Year%202023%2FMar2023.one%7CAC1235D8-7901-4ADA-996F-59FC28EE490E%2FCheck%20eq%3A110h%201%7CFC01D369-DFE1-8943-B275-F866B03F0B38%2F%29
onenote:https://uniroma3-my.sharepoint.com/personal/gitomassetti_os_uniroma3_it/Documents/Notebooks/Journal/0%20Year%202023/Mar2023.one#Check%20eq110h%201&section-id={AC1235D8-7901-4ADA-996F-59FC28EE490E}&page-id={FC01D369-DFE1-8943-B275-F866B03F0B38}&end}
\mynote{\myurl}
\begin{align}
	&\Big(\frac h R\mu\Big(1+\frac 13 \Big(\frac h R\Big)^2\Big)+\frac {p}2\Big)\Big(R\sP\Delta\sv+2\frac\sv R\Big)\nonumber\\
	&\qquad +\frac h R(\mu+\tilde\lambda)\Big(1+\frac 1 3\Big(\frac h R\Big)^2\Big)\nabla\Big(R{\rm div}\sv+2 w\Big)\nonumber\\
	&\qquad -\frac 1 6\Big(\frac h R\Big)^3(2\mu+\widetilde\lambda)\nabla(R^2\Delta w+2w)
	=\varrho h\sv'',\OK\nonumber\\
	&\frac 1 {12}\Big(\frac h R\Big)^3\Big[(2\mu+\widetilde\lambda)(-R^3\Delta\Delta w)+2\mu R^2\operatorname{div}(\sP\Delta\sv)\nonumber\\
&\qquad \qquad\, +2(\mu+\widetilde\lambda)R^2\Delta\operatorname{div}\sv-4(2\mu+\widetilde\lambda)\frac w R\Big]\nonumber\\
	&\qquad+\Big(\frac p 2+\frac 1 6 \Big(\frac h R\Big)^3
(\mu+2\widetilde\lambda)\Big)\Big(R\Delta w+2\frac w R\Big)\nonumber\\
&\qquad-\Big(2\frac h R(\mu+\tilde\lambda)+\frac 1 3 \Big(\frac h R\Big)^3\widetilde\lambda\Big)\Big({\rm div}\sv+2\frac w R\Big)=\varrho h w''.\OK\label{eq:110h}
	\end{align}
By performing the same \color{black}adimensionalisation that \color{black}led to \eqref{eq:99h}, we arrive at
\begin{align}
	&\Big(1+\frac 1 2 \frac {P}{H}+\frac 1 3 H^2\Big)(\sP\Delta\sv+2\sv)+(1+\Lambda)\Big(1+\frac 1 3 H^2\Big)\nabla({\rm div}\sv+2 w)\nonumber\\
	&\qquad\qquad -\frac 1 6 H^2(2+\Lambda)\nabla(\Delta w+2w)=\sv'',\OK\nonumber\\
	&
	\frac 1 {12}H^2\big((2+\Lambda)(-\Delta\Delta w+2\operatorname{div}(\sP\Delta\sv)+2(1+\Lambda)\Delta\operatorname{div}\sv-4(2+\Lambda)w\big)\nonumber\\
	&\qquad\qquad +\Big(\frac 12 \frac P H+\frac 1 6 H^2(1+2\Lambda)\Big)(\Delta w+2w)\nonumber\\
	&\qquad\qquad-\Big(2(1+\Lambda)+\frac 1 3 H^2\Lambda \Big)({\rm div}\sv+2w)=w''.\OK\label{eq:102}
	\end{align}
Instead of \eqref{eq:155yyy} we now have
\begin{equation}
\frac{{\rm d}^2}{{\rm d}t^2}\mathbbm u_n(t)=(\mathbb A_n+H^2\mathbb B_n)\mathbbm u_n(t),
\end{equation}
where $\mathbb A_n$ is the matrix defined in \eqref{eq:146ppp}, and where
\begin{equation}
\mathbb B_n=\begin{pmatrix}
\displaystyle\frac 1 3 (2(1-s_n^2)-s_n^2\Lambda)&\displaystyle\frac {s_n} 6(2 s_n^2+(2+s_n^2)\Lambda)
\\[0.7em]
\displaystyle\frac {s_n} 6(2 s_n^2+(2+s_n^2)\Lambda) &\displaystyle	-\frac 1 {12}(2(2+s_n^2+s_n^4)+(2+s_n^2)^2)\Lambda.
\end{pmatrix}.
\end{equation}
\color{black}For \color{black}a spherical shell to be considered ``thin'', the thickness $h$ should be at least one order of magnitude smaller than the radius $R$, which means that the correction associated to bending terms is two orders of magnitude smaller than that from the membrane terms. This is in agreement with \cite{kuoSmallOscillationsPressurized2015}, where it has been observed that the contribution from bending moments is negigible.

\section*{Acknowledgements}
This work was supported by the Istituto Nazionale di Alta Matematica (INdAM) -- Gruppo Nazionale per la Fisica Matematica (GNFM), and by MUR through PRIN Project DISCOVER (CUP: F53D2300184). 

\bibliographystyle{elsarticle-harv}
\bibliography{2023shells.bib}

\section*{Appendix}

\subsection*{A.1 Toolbox}\label{app:divergence}
Given a smooth tensor field $\bv$ on $\mathcal S$, we define the gradient of $\bv$ as the unique superficial tensor field $\nabla\bv$ such that, for every $\bfx\in\mathcal S$, 
  \begin{equation}
    \nabla\bv(\mathbf x)\boldsymbol\gamma'(0)=\left.\frac{{\rm d}}{{\rm d}s}\right|_{s=0}\bv(\boldsymbol\gamma(s))
  \end{equation}
for every smooth curve $\boldsymbol\gamma:\mathbb R\to\mathcal S$ such that $\boldsymbol\gamma(0)=\bfx$. Notice that, in particular
\begin{equation}\label{eq:174bbb}
	\nabla\bfx=\sP,
\end{equation}
where $\sP(\bfx)=\bI-\bn(\bfx)\otimes\bn(\bfx)$ is the orthogonal projection on the tangent space $T_\bfx\mathcal S$.

If $\sv$ is a tangential vector field, we define the divergence of $\sv$ as
  \begin{equation}\label{eq:70b}
    {\rm div}\sv=\sP\cdot\nabla\sv.
  \end{equation}
  Given a smooth part $\mathcal P$ of $\mathcal S$, let $\sn_{\mathcal P}$ be the tangent unit vector on $\partial\mathcal P$ that points away from the interior of $\mathcal P$.
\begin{figure}[h] \centering
\begin{tikzpicture}
\node[draw=none,fill=none] at (0,0){\includegraphics[width=0.3\paperwidth]{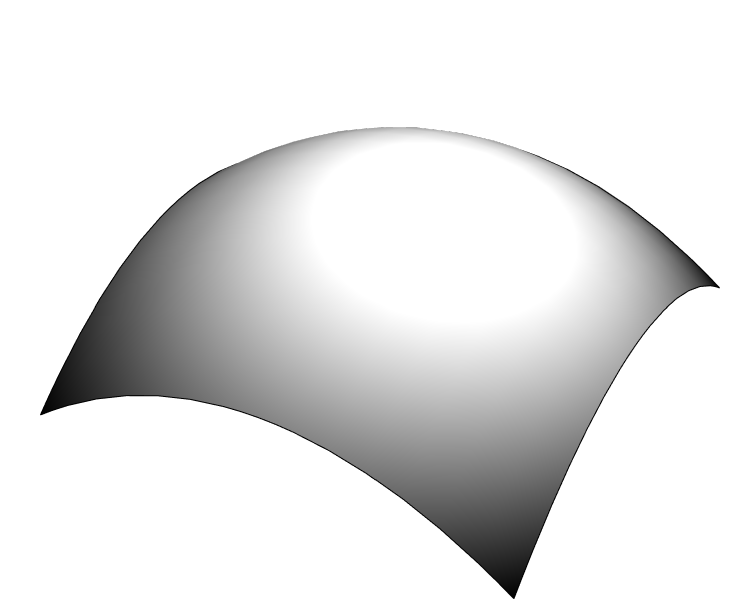}};
\draw (1,-1.5) node {$\mathcal S$};
\draw (0,0.5) node {$\mathcal P$};
\draw [very thick] (-1,1) 
      to [out=20,in=140] (1.5,0.5) 
      to [out=-40,in=0] (0,-.5) 
      to [out=180,in=200] (-1,1);
\draw (1,-1.5) node {$\mathcal S$};
\draw [thin,color=blue,->](0.8,0.45) -- +(0.5,-0.4);
\draw [thin,color=blue,->](0,0) -- +(0.4,-0.2);

\draw [thin,color=blue,->](-1,1.1) -- +(0.5,0);
\draw [thin,color=blue,->](-0.3,1) -- +(0.4,-0.1) ;
\draw [thin,color=blue,->](0.2,0.8) -- +(0.4,-0.2);
\draw [thin,color=blue,->](-1.3,0.5) -- +(0.5,-0.1);
\draw [thin,color=blue,->](-0.6,0.3) -- +(0.4,-0.15) ;
\draw [thin,color=blue,->](0.6,-0.3) -- +(0.4,-0.3);
\draw [thin,color=blue,->](0.3,-0.7) -- +(0.4,-0.2);
+(0.4,-0.1) ;
\draw [thin,color=blue,->](-0.8,-0.4) -- +(0.4,-0.1) ;
\draw [thin,color=blue,->](-1.6,-0.2) -- +(0.4,-0.05);
\draw [thin,color=blue,->](0.4,1.3) -- +(0.4,-0.1);
\draw [thin,color=blue,->](1.2,1) -- +(0.4,-0.2) node [above] {$\color{black}\mathsf v$};
\draw [very thick,-stealth](1.55,0.1) -- +(0.5,-0.2) node [above]{$\boldsymbol{\mathsf n}_{\mathcal P}$};
\draw [very thick,-stealth](1,-0.3) -- +(0.3,-0.4) node [above]{};
\draw [very thick,-stealth](0,-0.5) -- +(0.01,-0.4) node [above]{};
\draw [very thick,-stealth](-1,0.1) -- +(-0.4,-0.1) node [above]{};
\draw [very thick,-stealth](-1.2,0.8) -- +(-0.4,0.1) node [above]{};
\draw [very thick,-stealth](-0.3,1.15) -- +(-0.05,0.3) node [above]{};
\draw [very thick,-stealth](0.9,0.9) -- +(0.2,0.3) node [above]{};
\end{tikzpicture}
\caption{A surface $\mathcal S$ with a part $\mathcal P$, the field $\sn_{\mathcal P}$ perpendicular to $\partial\mathcal P$ (thick lines), and the tangential vector field $\sv$ (thin lines).}
\end{figure}
For $\color{black}\boldsymbol{\mathsf{v}}$ a smooth tangential vector field, we shall make use of the following identity:
\begin{equation}\label{eq:17b}
\int_{\partial \mathcal{P}} \sv \cdot \sn_{\mathcal P}=\int_{\mathcal{P}} \operatorname{div} \sv.
\end{equation}

To define the divergence of a tangential tensor field we first consider a tensor field of the form $\ba\otimes\sv$ where $\ba$ is a constant vector and $\sv$ is a tangential vector field, and we define:
  \begin{equation}\label{eq:8}
    {\rm div}(\ba\otimes\sv)=({\rm div}\sv)\ba
  \end{equation}
  For any such field, we have
    \begin{equation}\label{eq:86c}
 \int_{\partial\mathcal P}(\ba\otimes\sv)\sn_{\mathcal P}=\Big(\int_{\partial\mathcal P}\sv\cdot\sn_{\mathcal P}\Big)\ba=
 \Big(\int_{\mathcal P}{\rm div}\sv\Big)\ba=
 \int_{\mathcal P}({\rm div}\sv)\ba= \int_{\mathcal P}{\rm div}(\ba\otimes\sv).
  \end{equation}
  The above identity carries over to arbitrary tangential tensor fields, since each such field can be decomposed into a linear combination of dyadic products between constant vectors and tangential tensor fields. One way to obtain such decomposition is to fix an orthonormal basis $\mathbf e_i$, $i=1,2,3$ for $T\mathcal E$. Then $\bI=\sum_{i=1}^3\mathbf e_i\otimes\mathbf e_i$ and hence $\bT=\sum_{i=1}^3\mathbf e_i\otimes\mathbf T^\top\mathbf e_i=\sum_i\mathbf e_i\otimes\sv^{(i)}$
  with $\sv^{(i)}=\mathbf T^\top\mathbf e_i$.  
   Then, by linearity, we obtain
\begin{equation}
\int_{\partial\mathcal P}\bT\,\sn_{\mathcal P}=\int_{\mathcal P}{\rm div}\,\bT.	
\end{equation}
We observe from \eqref{eq:8} that $\operatorname{div}(\ba\otimes\sv)=(\nabla\sv\cdot\sP)\ba=\nabla(\ba\otimes\sv)\sP$ (here the third-order tensor $\nabla(\aa\otimes\sv)$ is regarded as a linear mapping from second-order tensors into vectors). Again, by linearity, we have
\begin{equation}
\operatorname{div}\bT=\nabla\bT:\sP,
\end{equation}
and, or every constant vector field $\ba$,
  \begin{equation}
  \ba\cdot{\rm div}\,\bT={\rm div}(\bT^\top\ba).
  \end{equation}
In particular, if $\sigma$ is a scalar field, then $\ba\cdot\operatorname{div}(\sigma\bT)=\operatorname{div}(\sigma\bT^\top\ba)=\nabla\sigma\cdot(\bT^\top\ba)+\sigma\operatorname{div}(\bT^\top\ba)=\ba\cdot(\bT\nabla\sigma+\sigma\operatorname{div}\bT)$. Hence,
\begin{equation}\label{eq:181bbb}
\operatorname{div}(\sigma\bT)=\bT\nabla\sigma+\sigma\operatorname{div}\bT.	
\end{equation}
If $\bv$ is a vector field, then $\operatorname{div}(\bT^\top\bv)=\nabla(\bT^\top\bv):\sP=\nabla\bT:(\bv\otimes\sP)+(\bT^\top\nabla\bv):\sP$, whence
\begin{equation}
\operatorname{div}(\bT^\top\bv)=	\operatorname{div}\bT\cdot\bv+\bT:\nabla\bv.
\end{equation}
Given two second-order tensor fields $\bA$ and $\boldsymbol B$, and a constant vector $\ba$, we have $\ba\cdot\operatorname{div}(\bA\boldsymbol B)=\operatorname{div}(\boldsymbol{B}^\top\bA^\top\ba)=\boldsymbol B:\nabla(\bA^\top\ba)+\bA^\top\ba\cdot\operatorname{div}\boldsymbol{B}=\nabla\bA:(\ba\otimes\boldsymbol B)+\ba\cdot(\bA\operatorname{div}\boldsymbol{B})=\ba\cdot(\nabla\bA:\boldsymbol B+\bA\operatorname{div}\boldsymbol{B})$, whence
\begin{equation}\label{eq:182bbb}
\operatorname{div}(\bA\boldsymbol B)=\nabla\bA:\boldsymbol B+\bA\operatorname{div}\boldsymbol{B}.
\end{equation}
The previous definitions and the identities can easily be generalized to tensor fields of arbitrary order. In particular, if $\mathbb T$ is a third-order tensor field, then, regarding $\mathbb T$ at each point as a linear map of vectors into second-order tensors, we define the divergence of $\mathbb T$ by requiring that, for every constant second-order tensor $\mathbf A$,
\begin{equation}
\mathbf A:{\rm div}\mathbb T	={\rm div}(\mathbf A:\mathbb T),
\end{equation}
where $\mathbf A:\mathbb T$ is the unique vector field $\sv$ such that $(\mathbf A:\mathbb T)\cdot\sa=\mathbf A\cdot(\mathbb T\sa)$ for every constant vector $\sa$. 
Consider in particular the third-order tensor field $\mathbb T=\bv\otimes\bT$ where $\bv$ is a vector field and $\bT$ is a tangential tensor field. Then
\begin{equation}\label{eq:90d}
	{\rm div}(\bv\otimes\bT)=\nabla\bv\,\bT^\top+\boldsymbol\bv\otimes{\rm div}\,\bT.
\end{equation}

\subsection*{A.2 Proof of the part-wise equilibrium equations \eqref{eq:44}}\label{app:partwise}
Starting from \eqref{eq:1}, we have
\begin{equation}
\int_{\partial\mathcal P}\bS\sn_{\mathcal P}\wedge(\bfy-\bfo)=-2{\rm skw}\int_{\partial\mathcal P}((\bfy-\bfo)\otimes\bS)\sn_{\mathcal P}=-2{\rm skw}\int_{\mathcal P}{\rm div}((\bfy-\bfo).\otimes\bS)
\end{equation}
Using the identity \eqref{eq:90d}, we obtain
\begin{equation}
{\rm div}((\bfy-\bfo)\otimes\bS)=\nabla\bfy\bS^\top+(\bfy-\bfo)\otimes{\rm div}\bS.	
\end{equation}
Thus, 
\begin{equation}
	\begin{aligned}[b]
\int_{\partial\mathcal P}\bS\sn_{\mathcal P}\wedge(\bfy-\bfo)&=2{\rm skw}\int_{\mathcal P}	\big(\bS\nabla\bfy^\top+{\rm div}\bS\otimes(\bfy-\bfo)\big)\\
&=\int_{\mathcal  P}\big({\rm div}\bS\wedge(\bfy-\bfo)+2{\rm skw}(\bS\nabla\bfy^\top)\big).
\end{aligned}
\end{equation}
A similar calculation shows that
\begin{equation}
\int_{\partial\mathcal P}\bM\sn_{\mathcal P}\wedge\bd=
\int_{\mathcal  P}\big({\rm div}\bM\wedge\bd+2{\rm skw}(\bM\nabla\bd^\top)\big)	.
\end{equation}




\subsection*{A.3 Cordinate systems}\label{app:curvilinear}
To compare our results with the literature based on coordinates, we provide in this subsection some tools for the conversion of our coordinate-free notation into the coordinate-based notation. 

We assume that the surface $\mathcal S$ can be parametrized through a smooth map $\widehat\bfx:\Sigma\to\mathcal S$, where $\Sigma$ is a domain of $\mathbb R^2$. Thus, the typical point $\bfx$ on $\mathcal S$ can be written as 
\begin{equation}
	\bfx=\widehat\bfx(\zeta^1,\zeta^2).
\end{equation}
We require the function $\widehat\bfx$ to be invertible, and we denote by
\begin{equation}
\bfx\mapsto(\widetilde\zeta^1(\bfx),\widetilde\zeta^2(\bfx))	
\end{equation}
its inverse. Every field $w$ defined on $\mathcal S$ can be regarded both as a function of $\bfx$ and of the coordinates $(\zeta^1,\zeta^2)$: we use a superposed hat in the former case, and a superposed tilde in the latter. Thus, we write 
$w=\widetilde w(\bfx)=\widehat w(\zeta^1,\zeta^2)$. To simplify the notation, we will omit hats and tildes in formulas where the context allows for clear interpretation.

We introduce the covariant basis
\begin{equation}
\sa_\alpha=\frac{\partial\bfx}{\partial \zeta^\alpha}=\bfx_{,\alpha},\quad\alpha=1,2,
\end{equation}
and the physical basis $\sa_{\langle\alpha\rangle}=\sa_\alpha/|\sa_\alpha|$. If the coordinates $\zeta^\alpha$ are dimensionless, then
\begin{equation}
[\sa_\alpha]={\rm Length},\qquad [\sa_{\langle\alpha\rangle}]=1,	
\end{equation}
i.e., the vectors of the covariant basis have dimension of length, whereas the vectors of the physical basis are dimensionless. We observe that since $\sa_{\alpha,\beta}=\bfx_{,\alpha\beta}=\bfx_{,\beta\alpha}$ we have
\begin{equation}\label{eq:196hh}
	\sa_{\alpha,\beta}=\sa_{\beta,\alpha}.
\end{equation}
The unit normal $\bn$ is given by
\begin{equation}\label{eq:14a}
  \bn=a^{-1}\sa_1\times\sa_2,\qquad a=|\sa_1\times\sa_2|.
\end{equation}
We define the contravariant basis by
\begin{equation}
\sa^\alpha=\nabla\zeta^\alpha,\quad\alpha=1,2.	
\end{equation}
The contravariant basis is related to the covariant basis by the relation
\begin{equation}\label{eq:129h}
\begin{aligned}
	\sa^1=-\frac{\bn\times\sa_2}{\bn\cdot\sa_1\times\sa_2},\qquad
	\sa^2=\frac{\bn\times\sa_1}{\bn\cdot\sa_1\times\sa_2},
\end{aligned}	
\end{equation}
and has the property that
\begin{equation}\label{eq:199g}
\sa_\alpha\cdot\sa^\beta=\delta_\alpha^\beta.	
\end{equation}
Again, if the coordinates are dimensionless, then the elements of the contravariant basis have dimensions of reciprocal of a length:
\begin{equation}
[\sa^\alpha]={\rm Length}^{-1}.	
\end{equation}
By taking the surface gradient of both sides of the equation $\bfx=\widehat\bfx(\tilde\zeta^1(\bfx),\tilde\zeta^2(\bfx))$ and by noting that $\nabla\bfx=\sP$, we obtain, by the chain rule,
\begin{equation}\label{eq:130j}
\sP=\sa_\alpha\otimes\sa^\alpha.	
\end{equation}
Here and in what follows we use the Einstein convention for repeated indices, with Greek letters ranging from 1 to 2. In particular, it follows from \eqref{eq:130j} that if $\sv$ is a tangential field, then the components of $\sv$ in the covariant basis can be obtained by scalar multiplication with the elements of the covariant basis, viz.
\begin{equation}
\sv=\sP\sv=(\sv\cdot\sa^\alpha)\sa_\alpha=\mathsf v^\alpha\sa_\alpha,\qquad\qquad \mathsf v^\alpha=\sv\cdot\sa^\alpha. 	
\end{equation}
The quantities $\mathsf v^\alpha$ are called contravariant components of $\sv$, and should not confused with the physical components $v_{\langle\alpha\rangle}$, which appear in the representation:
\begin{equation}
\sv=v_{\langle\alpha\rangle}\sa_{\langle\alpha\rangle}.	
\end{equation}
A further application of the chain rule yields
\begin{equation}\label{eq:130h}
\nabla w=\sum_{\alpha=1,2}\frac{\partial w}{\partial\zeta^\alpha}\nabla\zeta^\alpha=w_{,\alpha}	\sa^\alpha.
\end{equation}
Likewise, for $\bv$ a vector field, 
\begin{equation}\label{eq:132h}
\nabla\bv=\bv_{,\alpha}\otimes\sa^\alpha.	
\end{equation}
Thus, if $\sv$ is a tangential vector field, then by \eqref{eq:70b} we have ${\rm div}\sv=\sP\cdot(\sv_{,\alpha}\otimes\sa^\alpha)=\sP\sv_{,\alpha}\cdot\sa^\alpha=\sv_{,\alpha}\cdot\sP\sa^\alpha$, since $\sP=\sP^\top$, and hence
\begin{equation}
\operatorname{div}\sv=\sv_{,\alpha}\cdot\sa^\alpha.	
\end{equation}
Furthermore, introducing the Christoffel symbols
\begin{equation}
	 \Gamma_{\alpha\gamma}^\beta=\sa_{\alpha,\gamma}\cdot\sa^\beta,
\end{equation}
we can write
\begin{equation}
	\boldsymbol{\mathsf v}_{,\alpha}=(v^\beta\boldsymbol{\mathsf a}_\beta)_{,\alpha}=(v^\beta_{,\alpha}+v^\gamma\Gamma^\beta_{\alpha\gamma})\sa_\beta.
\end{equation}
Therefore, on setting
\begin{equation}
	\mathsf v^\beta|_{\alpha}=\Gamma^\beta_{\alpha\gamma} v^\gamma,
\end{equation}
we obtain the following expression for the surface divergence:
\begin{equation}\label{eq:202}
\operatorname{div}\sv=v^\alpha|_\alpha.
\end{equation}
An alternative expression is
\begin{equation}\label{eq:203}
	\operatorname{div}\sv=\frac 1 {\sqrt a} ({\sqrt a} v^\alpha)_{,\alpha},
\end{equation}
where $a$ has been defined in \eqref{eq:14a}. To verify \eqref{eq:203}, we recall that $\sa_{\alpha,\beta}=\bfx_{,\alpha\beta}=
\sa_{\alpha,\beta}$, and we use \eqref{eq:129h}, we obtain
\begin{equation}
	\begin{aligned}[b]
	\frac 1 a (a v^\alpha)_{,\alpha}&=\frac 1 {\sqrt a} (\sa_{1,\alpha}\times\sa_2\cdot\bn) v^\alpha+\frac 1 {\sqrt a} (\sa_1\times\sa_{2,\alpha}\cdot\bn \,v^\alpha)+v^\alpha_{,\alpha}\\
	&=\frac 1 {\sqrt a} (\sa_{\alpha,1}\times\sa_2\cdot\bn) v^\alpha+\frac 1 {\sqrt a} (\sa_1\times\sa_{\alpha,2}\cdot\bn) v^\alpha+v^\alpha_{,\alpha}\\
	&=\Big(\sa_{\alpha,1}\cdot\frac{\sa_2\times\bn}{\sqrt a}\Big) v^\alpha+\Big(\sa_{\alpha,2}\cdot \frac{\bn\times\sa_1}{\sqrt a}\Big) v^\alpha+v^\alpha_{,\alpha}\\
		&=(\sa_{\alpha,1}\cdot\sa^1) v^\alpha+(\sa_{\alpha,2}\cdot \sa^2) v^\alpha+(\sa_\alpha\cdot\sa^\beta)v^\alpha_{,\beta}\\
			&=(\sa_{\alpha,\beta}\cdot\sa^\beta) v^\alpha+(\sa_\alpha\cdot\sa^\beta)v^\alpha_{,\beta}=(v^\alpha \sa_\alpha)_{,\beta}\cdot\sa^\beta\\
			&=\sv_{,\beta}\cdot\sa^\beta,
	\end{aligned}
\end{equation}
as required.

One can also check that if $\bT$ is a tensor field of any order, then 
\begin{equation}\label{eq:207b}
{\rm div}\bT=\bT_{,\alpha}\sa^\alpha.
\end{equation}
In particular, we can easily recover \eqref{eq:90d} by computing
\begin{equation}
{\rm div}(\bv\otimes\bT)=(\bv\otimes\bT)_{,\alpha}\otimes\sa^\alpha=\bv_{,\alpha}\otimes\bT\sa^\alpha+\bv\otimes{\rm div}\,\bT.
\end{equation}
and by observing that $\bv_{,\alpha}\otimes\bT\sa^\alpha=(\bv_{,\alpha}\otimes\sa^\alpha)\bT^\top=\nabla\bv\bT^\top$.

\subsection*{A.3 More on the pseudo-inverse $\bF^{-1}$.}\label{app:pseudo}
We begin by proving  that the requirements $\bF^{-\top}\bn=0$, $\bF\bF^{-1}=\overline\bP$ and $\bF^{-1}\bF=\sP$ determine a unique pseudo-inverse $\bF^{-1}$. Suppose indeed that there exists two tensors, say $\bF^{-1}$ and $\tilde\bF^{-1}$ which satisfy the above requirements. Let $\boldsymbol D=\bF^{-1}-\tilde\bF^{-1}$. Then $\boldsymbol D^{-\top}\bn=\mathbf 0$, $\boldsymbol D\bF=\mathbf 0$ and $\bF\boldsymbol D=\mathbf 0$. From the equation $\boldsymbol D\bF=\mathbf 0$ we deduce that the null-space of $\boldsymbol D(\bfx)$ is the range of $\bF(\bfx)$, i.e., the space $T_{\bfy(\bfx)}\overline{\mathcal S}$; this fact implies that $\boldsymbol D=\boldsymbol a\otimes\overline{\boldsymbol n}$ for some vector $\sa$. From the equation $\bF\boldsymbol D=\mathbf 0$ we deduce that $\sa$ belongs to the null space of $\bF$, and hence is parallel to the normal vector $\bn$. Accordingly, we have $\boldsymbol D=\alpha\bn\otimes\overline{\bn}$ for some scalar $\alpha$. However, since $\boldsymbol D^{\top}\bn=\mathbf 0$, the scalar $\alpha$ must necessarily vanish. Thus, we conclude that $\boldsymbol D=\mathbf 0$, and hence $\bF^{-1}$ is unique.

We next construct the pseudo-inverse at a given point $\bfx$ under the assumption that $\bd(\bfx)$ does not belong to the tanget space $T_{\bfy(\bfx)}\overline{\mathcal S}$. To this effect, we select two linearly-independent vectors $\sa_1$ and $\sa_2$ in the tangent space at $\bfx$ and we let $\overline\sa_\alpha=\bF\sa_\alpha$. We also define $\sa_3=\bn$ and $\overline\sa_3=\bd$. Next we let $(\sa^1,\sa^2,\sa^3)$, and $(\overline\sa^1,\overline\sa^2,\overline\sa^3)$ be the reciprocal bases of, respectively, $(\sa_1,\sa_2,\sa_3)$ and $(\overline\sa_1,\overline\sa_2,\overline \sa_3)$. We recall that the reciprocal bases are defined by the requirement that $\sa_i\cdot\sa^j=\overline\sa_i\cdot\overline\sa^j=\delta_i^j$, and can be constructed with the formulas $\sa^i=(\sa_j\times\sa_k)/(\ba_i\cdot(\ba_j\times\ba_k))$ for $i\neq j\neq k$. It is easy to check that $\ba^3=\bn$, and that $\overline\ba^3=(\bd\cdot\overline\bn)^{-1}\overline\bn$.
With these definitions, the deformation gradient can be written as $\bF=\overline\ba_\alpha\otimes\ba^\alpha$. Moverover, the projectors $\sP$ and $\overline\bP$ are given, respectively, by
\begin{equation}
\sP=\ba_\alpha\otimes\ba^\alpha\qquad\text{and}\qquad \overline\bP=\overline\ba_\alpha\otimes\overline\ba^\alpha.
\end{equation}
We conclude by checking that
\begin{equation}
\bF^{-1}=\ba_\alpha\otimes\overline\ba^\alpha
\end{equation}
is the required pseudo-inverse. Indeed, we have $\bF^{-\top}\bn=(\overline\ba^\alpha\otimes\ba_\alpha)\bn=(\overline\ba^\alpha\otimes\ba_\alpha)\ba^3=\mathbf 0$. Moreover, $\bF\bF^{-1}=(\overline\ba_\alpha\otimes\ba^\alpha)(\ba_\beta\otimes\overline\ba^\beta)=(\ba^\alpha\cdot\ba_\beta)(\overline\ba_\alpha\otimes\overline\ba^\beta)=\delta_\beta^\alpha(\overline\ba_\alpha\otimes\overline\ba^\beta)=\overline\ba_\alpha\otimes\overline\ba^\alpha=\overline\bP$. A similar calculation can be used to check that $\bF^{-1}\bF=\sP$.

\subsection*{A.4 Proof of the divergence identity \eqref{eq:17b}.}\label{app:diver_proof}
The following proof of the divergence identity is adapted from \cite{fried2021mechanics}. Let $\st_{\mathcal P}=\bn\times\sn_{\mathcal P}$. Then
\begin{equation}
	   \int_{\partial\mathcal P}\sv\cdot\sn_{\mathcal P}=\int_{\partial\mathcal P}\bn\times\sv\cdot\st_{\mathcal P}.
\end{equation}
Granted that $\mathcal S$ is smooth, there exists a neighborhood of $\mathcal S$ in which $\sv$ and $\bn$ admit normally constant extensions, which we still denote by $\sv$ and $\bn$, respectively. By Stokes' Theorem,
\begin{equation}
	\int_{\partial\mathcal P}{\bn\times\sv}\cdot\st_{\mathcal P}=\int_{\mathcal P}\operatorname{curl}(\bn\times\sv)\cdot\bn.
	\end{equation}
We now recall that  
\begin{equation}\label{eq:210uuu}
\operatorname{curl}(\bn\times\bv)\cdot\bn=\nabla(\bn\times\sv)\cdot (\bn\times),
\end{equation}
where $(\bn\times)$ is the unique skew-symmetric tensor such that $\bn\times\ba=(\bn\times)\ba$ for every vector $\ba$. 
Note that, since both $\sv$ and $\bn$ are constant along the normal direction, on $\mathcal S$ the gradient operator in \eqref{eq:210uuu} can be taken to be the surface gradient. Next, we write
\begin{equation}\label{eq:211uuu}
\nabla(\bn\times\sv)\cdot(\bn\times)=(\bn\times)\nabla\sv\cdot(\bn\times)-(\sv\times)\nabla\bn\cdot(\bn\times).
\end{equation}
As to the first term on the right-hand side of \eqref{eq:211uuu}, we observe that $(\bn\times)^2=-\sP$, and we write
\begin{equation}
	(\bn\times)\nabla\sv\cdot(\bn\times)=\nabla\sv\cdot(\bn\times)^\top(\bn\times)=-\nabla\sv\cdot(\bn\times)^2=\sP\cdot\nabla\sv=\operatorname{div}\sv.
\end{equation}
We now claim that the second term on the right-hand side of \eqref{eq:211uuu} vanishes. Indeed, since $\nabla\bn$ is symmetric, and $\bn\times$ is skew-symmetric,
\begin{equation}
	(\sv\times)\nabla\bn\cdot(\bn\times)=(\sv\times)\cdot(\bn\times)\nabla\bn=\frac 12 (\sv\times)\cdot\big((\bn\times)\nabla\bn-\nabla\bn(\bn\times)\big).
\end{equation}
Since $(\bn\times)\nabla\bn-\nabla\bn(\bn\times)$ is skew-symmetric, it is equal to $\ba\times$ for some vector $\ba$. On the other hand, $((\bn\times)\nabla\bn-\nabla\bn(\bn\times)\big)\bn=0$, which implies that $\ba$ is parallel to $\bn$. Thus, we conclude that $(\sv\times)\nabla\bn\cdot(\bn\times)=\frac 12(\sv\times)\cdot(\ba\times)=\sv\cdot\ba=0$, as desired.

\subsection*{A.5 Spherical coordinates}\label{sec:spherical}
To compare \eqref{eq:98h} with existing results, we introduce  a coordinate system on a sphere of radius $R$ through the parametrization
\begin{equation}
\widehat{\bfx}(\theta,\phi)=\boldsymbol o+R\sin \theta(\cos \phi \be_1+\sin \phi \be_2)+R\cos \theta \be_3,
\end{equation}
where $\boldsymbol e_i$, $i=1,\dots,3$ is an orthonormal basis, as shown in Fig.~\ref{fig:88}. 
\begin{figure}[h]
\centering
\begin{tikzpicture}[scale = 3]
 
\coordinate (P) at ({1/sqrt(3)},{1/sqrt(3)},{1/sqrt(3)});
\coordinate (O) at (0,0,0);
\coordinate (V) at (1,0,0);
\coordinate (R) at (0,1,0);
 
\shade[ball color = white,
    opacity = 0.1
] (0,0,0) circle (1cm);

\draw[dashed] (0,0,0) -- (0.8,0,0);
\draw[dashed] (0,0,0) -- (0,0.8,0);
\draw[dashed] (0,0,0) -- (0,0,1.30);

\draw[-stealth,thick] (0,0,0) -- (0,0.3,0) node [left] {$\be_3$};
\draw[-stealth,thick] (0,0,0) -- (0.3,0,0) node [above] {$\be_2$};
\draw[-stealth,thick] (0,0,0) -- (0,0,0.5) node [above] {$\be_1$};

\tdplotsetmaincoords{70}{110}
\tdplotdrawarc{(0,0,0)}{0.78}{0}{49}{anchor=north}{$\phi$}
\tdplotsetthetaplanecoords{60}
\tdplotdrawarc[tdplot_rotated_coords]{(0,0,0)}{0.5}{0}{45}{anchor=south west}{$\theta$}

\coordinate (Q) at (0.6,0,0.6);
\draw[thick, -stealth] (0,0,0) -- (P);
\draw [dashed] (P) -- (Q);
\draw [dashed] (0,0,0) -- (Q);

\draw[fill = lightgray!50] (P) circle (0.5pt);
 
 \draw [very thick,-stealth] (P) -- +(0.3,0.3,0.3) node [above] {$\bn$};
  \draw [very thick,-stealth] (P) -- +(0.3,-0.3,0.3) node [above right] {$\ba_{\langle\theta\rangle}$};
   \draw [very thick,-stealth] (P) -- +(0.5,0.3,0.3) node [above right] {$\ba_{\langle\phi\rangle}$};
 
 \draw (P)+(0.05,0.05,0.05) node [below right] {$\bfx$};
 \draw (0,0,0) node [below] {$\bfo$};
 
\end{tikzpicture}
\caption{The spherical coordinates $(\theta,\phi)$ and the associated physical basis $(\ba_{\langle\theta\rangle},\ba_{\langle\phi\rangle})$.}\label{fig:88}
\end{figure}
Since
\begin{equation}
	\bn(\bfx)=\frac1 R {(\bfx-\bfo)},
\end{equation}
a and since $\nabla\bfx=\sP$ (\emph{cf.} \eqref{eq:174bbb}), we have
\begin{equation}\label{eq:227bbb}
	\nabla\bn=\frac 1 R\sP,
\end{equation}
and hence
\begin{equation}\label{eq:228bbb}
\operatorname{div}\bn=\frac 2 R.	
\end{equation}
The covariant basis is
\begin{equation}\label{eq:covariant}
\begin{aligned}
&\sa_\theta=\bfx_{,\theta}=	R\cos\theta(\cos\phi\be_1+\sin\phi\be_2)+R\sin\theta\be_3,
\\
&\sa_\phi=\bfx_{,\phi}=R\sin\theta(-\sin\theta\be_1+\cos\theta\be_2).
\end{aligned}
\end{equation}
The normal vector is
\begin{equation}
\bn=\frac {\sa_\theta\times\sa_\phi} {|\sa_\theta\times\sa_\phi|	}=\frac{\bfx-\bfo}R=\sin\theta(\cos\theta\be_1+\sin\theta\be_2)+\cos\theta\be_3.
\end{equation}
According to \eqref{eq:129h} and \eqref{eq:covariant}, the contravariant basis is
\begin{equation}
	\begin{aligned}
	&\sa^\theta=
	\frac{\sa_\theta}{|\sa_\theta|^2}=\frac{\sa_\theta}R,
	\qquad
	\sa^\phi=\frac{\sa_\phi}{|\sa_\phi|^2}=
		\frac{\sa_\phi}{R^2\sin^2\theta}.
	\end{aligned}
\end{equation}
\subsection{Formulas in the covariant basis.} 
Consider scalar field $w$. On applying \eqref{eq:130h}, we write 
\begin{equation}\label{eq:161j}
\nabla w=w_{,\theta}\sa^\theta+w_{,\phi}\sa^\phi.
\end{equation}
When applying differential operators, the resulting computations necessitate taking the derivatives of the covariant basis vectors. For ease and clarity, it's advantageous to represent these derivatives as linear combinations of the covariant basis and $\bn$. An important aspect to remember during these calculations is that $\sa_\alpha$, being the partial derivative of $\bfx$ with respect to $\alpha$, obeys the symmetry condition $\sa_{\alpha,\beta}=\sa_{\beta,\alpha}$. This leads to 
\begin{equation}
\begin{aligned}
&\sa_{\theta,\theta}\cdot\sa^\theta=0,\qquad
&&\sa_{\theta,\theta}\cdot\sa^\phi=0,\\
&\sa_{\theta,\phi}\cdot\sa^\theta=\sa_{\phi,\theta}\cdot\sa^\theta=0,\qquad
&&\sa_{\theta,\phi}\cdot\sa^\phi=\sa_{\phi,\theta}\cdot\sa^\phi=\cot(\theta),\\
&\sa_{\phi,\phi}\cdot\sa^\theta=-\cos(\theta)\sin(\theta),\quad
&&\sa_{\phi,\phi}\cdot\sa^\phi=0,
\end{aligned}
\end{equation}
and
\begin{equation}
	\begin{aligned}
		&\sa_{\theta,\theta}\cdot\bn=-R,\\
		&\sa_{\theta,\phi}\cdot\bn=\sa_{\phi,\theta}\cdot\bn=0,\\
	&\sa_{\phi,\phi}\cdot\bn=-R\sin^2(\theta).
	\end{aligned}
\end{equation}
Thus, we can write
\begin{equation}
	\begin{aligned}
	&\sa_{\theta,\theta}=-R\bn,\qquad
	&\sa_{\theta,\phi}=\sa_{\phi,\theta}=\cot(\theta)\sa_\phi,
		\end{aligned}
\end{equation}
and
\begin{equation}
	\begin{aligned}
		&\sa_{\phi,\phi}=-\cos(\theta)\sin(\theta)\,\sa_\theta-R\sin^2(\theta)\,\bn.
	\end{aligned}
\end{equation}
As a consequence, we have
\begin{equation}
\begin{aligned}
&\sP\nabla \sa_\alpha=\sa_{\alpha,\beta}\otimes\sa^\beta=(\sa_{\alpha,\beta}\cdot\sa^\gamma)\sa_\gamma\otimes\sa^\beta,\\
&{\rm div}\sa_\alpha=\sa_{\alpha,\beta}\cdot\sa^\beta.
\end{aligned}
\end{equation}
In particular
\begin{equation}
\begin{aligned}
&\sP\nabla\sa_\theta	=\cot(\theta)\sa_\phi\otimes\sa^\phi,&&{\rm div}\sa_\theta=\cot(\theta),\\
&\sP\nabla\sa_\phi=\cot(\theta)\sa_\phi\otimes\sa^\theta-\cos(\theta)\sin(\theta)\sa_\phi\otimes\sa^\phi,&&{\rm div}\sa_\phi=-\cos(\theta)\sin(\theta),
\end{aligned}
\end{equation}
Consider now a tangential vector field
\begin{equation}
\sv=v^\alpha	\sa_\alpha.
\end{equation}
Using \eqref{eq:132h} we find 
\begin{equation}\label{eq:166}
\begin{aligned}[b]\sP\nabla\sv&=(v^\alpha\sa_\alpha),_\beta\otimes\sa^\beta=v^\alpha_{,\beta}\sa_\alpha\otimes\sa^\beta+v^\alpha\sa_{\alpha,\beta}\otimes\sa^\beta\\
&=v^\alpha_{,\beta}\sa_\alpha\otimes\sa^\beta+v^\alpha\sa_{\alpha,\beta}\otimes\sa^\beta\\
&=v^\alpha_{,\beta}\sa_\alpha\otimes\sa^\beta+v^\gamma(\sa_{\gamma,\beta}\cdot\sa^\alpha)\sa_\alpha\otimes\sa^\beta.
\end{aligned}	
\end{equation}
 This, in turn, yields that the covariant derivative of $\sv$ is
\urldef\myurl\url{https://uniroma3-my.sharepoint.com/personal/gitomassetti_os_uniroma3_it/_layouts/OneNote.aspx?id=%2Fpersonal%2Fgitomassetti_os_uniroma3_it%2FDocuments%2FNotebooks%2FJournal&wd=target%280%20Year%202023%2FJan%202023.one%7CA3D34BC6-B674-4914-9451-C041DDCC17FD%2FComputation%20of%20grad%20v%7C5F2A4577-6705-D749-B54C-897944872A34%2F%29}
\mynote{\myurl}
\begin{equation}\label{eq:156h}
\begin{aligned}[b]
\sP\nabla\sv&=v^\theta_{,\theta}\sa_\theta\otimes\sa^\theta+(v^\theta_{,\phi}-(\cos\theta\sin\theta) \,v^{\phi})\sa_\theta\otimes\sa^\phi\\
&\quad+(v^{\phi}_{,\theta}+\cot(\theta)v^\phi)\sa_\phi\otimes\sa^\theta+(v^\phi_{,\phi}
+\cot(\theta)v^\theta)\sa_\phi\otimes\sa^\phi.
\end{aligned}
\end{equation}
Thus, the divergence of $\sv$ is
\begin{equation}
{\rm div}\sv=\sP\cdot\nabla\sv=v^{\theta}_{,\theta}+	v^{\phi}_{,\phi}+\cot(\theta)v^\theta.
\end{equation}
In particular, taking $\sv=\nabla w$ we find
\begin{equation}
\Delta w=\frac 1 {R^2}(w_{,\theta\theta}+\cot(\theta)w_{,\theta}	+\csc^2(\theta)w_{,\phi\phi}).
\end{equation}
By taking the divergence of \eqref{eq:166} we find
\begin{multline}
\sP{\rm div}\sP\nabla\sv=\frac 1{R^2}(v^\theta_{,\theta\theta}+\cot(\theta)(-\cot(\theta)v^\theta+v^\theta_{,\theta}-2v^\phi_{,\phi})+\csc^2(\theta)\,v^\theta_{,\theta\theta})\sa_\theta	\\
+\frac 1 {R^2}(-v^\phi+3\cot(\theta)v^\phi_{,\theta}+v^\phi_{,\theta\theta}+\csc^2(\theta)(2\cot(\theta)v^\phi_{\phi}+v^\phi_{\phi\phi}))\sa_\phi.
\end{multline}
Likewise, we obtain
\begin{multline}
\sP\Delta\sv=\frac 1 {R^2}(v^\theta_{,\theta\theta}+\cot(\theta)(v^\theta_{,\theta}-2v^\phi_{,\phi})+\csc^2(\theta)(-v^\theta+v^\theta_{,\phi\phi})\sa_\theta\\
+\frac 1 {R^2}(v^\theta_{,\theta\theta}+\cot(\theta)(v^\theta_{,\theta}-2v^\phi_{,\phi})+\csc^2(\theta)(v^\theta_{,\phi\phi}-v^\theta))\sa_\phi.
\end{multline}

\subsection{Formulas in terms of physical components.}
While the covariant and contravariant bases are especially useful for computational purposes, it's important to note that the components of a vectorial or tensorial quantity in these bases do not possess the same dimensions as the quantity in question. Consequently, it is sometimes advantageous to employ the physical basis, which is defined as follows:
\begin{equation}
	\begin{aligned}
	&\sa_{\langle\theta\rangle}=\frac {\sa_{\theta}} {|\sa_{\theta}|}=\frac {\sa_\theta} R,
,\qquad \sa_{\langle\phi\rangle}=\frac {\sa_{\phi}} {|\sa_{\phi}|}
	=\frac {\sa_\phi} {R\sin\theta}.
	\end{aligned}
\end{equation}
The physical basis is related to the contravariant basis by
\begin{equation}
	\begin{aligned}
	&\sa^\theta=
	\frac{\sa_{\langle\theta\rangle}}{R}
	, \qquad \sa^\phi=\frac{\sa_{\langle\phi\rangle}}{R\sin\theta}.
	\end{aligned}
\end{equation}
To ease the comparison with standard formulas, we use below the $\partial$ notation to denote partial derivatives. From \eqref{eq:161j} we have
\begin{equation}
\nabla w=\frac{\partial w}{\partial\theta}\sa^\theta+\frac{\partial w}{\partial\phi}\sa^\phi=\frac 1 R\frac{\partial w}{\partial\theta}\sa_{\langle\theta\rangle}+\frac{1}{R\sin(\theta)}\frac{\partial w}{\partial\phi}\sa_{\langle\phi\rangle}.
\end{equation}
Next, given a tangential vector field $\sv$, we introduce the physical components $v_{\langle\theta\rangle}$ and $v_{\langle\phi\rangle}$ of $\sv$ by writing
\begin{equation}
\sv=u_{\langle\theta\rangle}\sa_{\langle\theta\rangle}+u_{\langle\phi\rangle}\sa_{\langle\phi\rangle}.
\end{equation}
The physical components of $\sv$, are related to the contravariant components $v^\alpha$ by $v_{\langle\theta\rangle}=\sv\cdot\sa_{\langle\theta\rangle}=v^\theta\sa_\theta\cdot\sa_{\langle\theta\rangle}=Rv^\theta$ and 
$v_{\langle\phi\rangle}=\sv\cdot\sa_{\langle\phi\rangle}=v^\phi\sa_\phi\cdot\sa_{\langle\phi\rangle}=R\sin\theta v^\phi$.
Then \eqref{eq:156h} takes the form
\begin{equation}\label{eq:156h}
\begin{aligned}[b]
\sP\nabla\sv&=
\frac 1 R \frac{\partial v_{\langle\theta\rangle}}{\partial\theta}\sa_{\langle\theta\rangle}\otimes\sa_{\langle\theta\rangle}
+\frac 1 {R\sin\theta}\Big(\frac{\partial u_{\langle\theta\rangle}}{\partial\theta}-\cos\theta\frac{\partial u_{\langle\phi\rangle}}{\partial\phi}\Big)\sa_{\langle\phi\rangle}\otimes\sa_{\langle\theta\rangle}\\
&\quad
+
\frac 1 {R}\Big(\frac{\partial v_{\langle\phi\rangle}}{\partial\phi}+\frac{\cos\theta}{\sin\theta}v_{\langle\phi\rangle}\Big)\sa_{\langle\phi\rangle}\otimes\sa_{\langle\theta\rangle}
\\&\quad+\frac {1}{R\sin\theta}\Big(\frac{\partial v_{\langle\phi\rangle}}{\partial\phi}
+{\cos\theta}\,v_{\langle\theta\rangle}\Big)\sa_{\langle\phi\rangle}\otimes\sa_{\langle\phi\rangle}.
\end{aligned}
\end{equation}
The divergence of $\sv$ is then
\begin{equation}
{\rm div}\sv=\frac 1 R \frac{\partial v_{\langle\theta\rangle}}{\partial\theta}+
\frac {1}{R\sin\theta}\Big(\frac{\partial v_{\langle\phi\rangle}}{\partial\phi}
+{\cos\theta}\,v_{\langle\theta\rangle}\Big).
\end{equation}
We finally compute
\begin{equation}
\begin{aligned}[b]
\!\!\sP{\rm div}\sP\nabla\sv=&	
\frac 1 {R^2}\Big[\frac{\partial v_{\langle\theta\rangle}}{\partial\theta^2}+\frac{\cos(\theta)}{\sin(\theta)}\Big(-\frac{\cos(\theta)}{\sin(\theta)}v_{\langle\theta\rangle}+\frac{\partial v_{\langle\theta\rangle}}{\partial\theta}-\frac2 {\sin^2(\theta)}\frac{\partial v_{\langle\theta\rangle}}{\partial\phi}\Big)
\\
&\qquad+\frac 1 {\sin^2(\theta)}\frac{\partial^2 v_{\langle\theta\rangle}}{\partial\phi^2}\Big]\sa_{\langle\theta\rangle}\\
&+\frac 1 {R^2}\Big[-\frac{\cos^2(\theta)}{\sin^2(\theta)}v_{\langle\phi\rangle}+\frac{\cos(\theta)}{\sin(\theta)}\frac{\partial v_{\langle\phi\rangle}}{\partial\theta}\\
&\, \quad\qquad+\frac 1 {\sin^2(\theta)}\Big(2\cos(\theta)\frac{v_{\langle\theta\rangle}}{\partial\phi}+\frac{\partial^2v_{\langle\phi\rangle}}{\partial\phi^2}\Big)\Big]\sa_{\langle\phi\rangle},
\end{aligned}
\end{equation}
and
	\begin{align}
		\sP\Delta\sv=&\frac 1 {R^2}
		\Big[
			\frac{\partial^2 v_{\langle\theta\rangle}}{\partial\theta^2}
			+\frac{\cos(\theta)}{\sin(\theta)}\frac{\partial v_{\langle\theta\rangle}}{\partial\theta}
			\nonumber\\
			&\qquad -\frac 1 {\sin^2(\theta)}
			\Big(
				\frac{\partial^2 v_{\langle\theta\rangle}}{\partial\phi^2}
				+2\cos(\theta)\frac{\partial v_{\langle\phi\rangle}}{\partial\phi}
				+v_{\langle\theta\rangle}
			\Big)
		\Big]\sa_{\langle\theta\rangle}
		\nonumber\\
		&+\frac 1 {R^2}\Big[
			\frac{\partial^2 v_{\langle\phi\rangle}}{\partial\theta^2}
			+
			\frac{\cos(\theta)}{\sin(\theta)}\frac{\partial v_{\langle\phi\rangle}}{\partial\theta}
			\nonumber\\&\quad\qquad
			+\frac 1 {\sin^2(\theta)}
				\Big(
					\frac{\partial^2 v_{\langle\phi\rangle}}{\partial\phi^2}
				+2\cos(\theta)\frac{\partial v_{\langle\theta\rangle}}{\partial\phi}
				-v_{\langle\phi\rangle}
				\Big)
		\Big]\sa_{\langle\phi\rangle}.
	\end{align}
\subsection*{A.6 Proof of the identities \textbf{(\ref{eq:125})}}\label{app:identities}
\paragraph{Proof of $\bf(\ref{eq:125})_1$} We write
\begin{equation}
	\begin{aligned}[b]
	\sP\operatorname{div}(\sP\nabla\sv)&=\sP\operatorname{div}(\nabla\sv)-\sP\operatorname{div}(\bn\otimes\bn\nabla\sv)\\&=\sP\Delta\sv-\sP\operatorname{div}(\bn\otimes\nabla\sv^\top\bn).
	\end{aligned}
\end{equation} 
Then, since $\bn\cdot\sv=0$ and $\nabla\bn=\sP/R$, we obtain $\nabla\sv^\top\bn=\nabla(\bn\cdot\sv)-\nabla\bn^\top\sv=\sv/R$. Thus, we get $\sP\operatorname{div}(\bn\otimes\nabla\sv^\top\bn)\operatorname{div}(\bn\otimes\sv)=\bn\operatorname{div}\sv+\nabla\bn\sv=\sv/R$, as desired.

\paragraph{Proof of $\bf(\ref{eq:125})_1$} We use \eqref{eq:207b} and \eqref{eq:132h} to write
\begin{equation}
\begin{aligned}[b]
\sP\operatorname{div}(\nabla\sv^\top\sP)&=\sP(\nabla\sv^\top)_{,\alpha}\sa^\alpha=\sP(\sa^\beta\otimes\sv_{,\beta})_{,\alpha}\sa^\alpha+\nabla\sv^\top\operatorname{div}\sP\\
&=(\sv_{,\beta\alpha}\cdot\sa^\alpha)\sa^\beta+(\sv_{,\beta}\cdot\sa^\alpha)\sP\sa^\beta_{,\alpha}-(\operatorname{div}\bn)\nabla\sv^\top\bn,
\end{aligned}
\end{equation}
where $\operatorname{div}\sP=-\operatorname{div}(\bn\otimes\bn)=-(\operatorname{div}\bn)\bn-\nabla\bn\bn=-(\operatorname{div}\bn)\bn$.
Then, we notice that
\begin{equation}
\begin{aligned}[b]
\nabla(\operatorname{div}\sv)&=(\sv_{,\beta}\cdot\sa^{\beta})_{,\alpha}\sa^\alpha=(\sv_{,\beta\alpha}\cdot\sa^\beta)\sa^\alpha+(\sv_{,\beta}\cdot\sa^\beta_{,\alpha})\sa^\alpha\\
&=(\sv_{,\beta\alpha}\cdot\sa^\alpha)\sa^\beta+(\sv_{,\beta}\cdot\sa^\alpha_{,\beta})\sa^\beta.
\end{aligned}
\end{equation}
Thus,
\begin{equation}\label{eq:251bbb}
	\sP\operatorname{div}(\sP\nabla\sv^\top)-\nabla(\operatorname{div}\sv)=(\sv_{,\beta}\cdot\sa^\alpha)\sP\sa^\beta_{,\alpha}-(\sv_{,\beta}\cdot\sa^\alpha_{,\beta})\sa^\beta-(\operatorname{div}\bn)\nabla\sv^\top\bn.
\end{equation}
Now, recalling that $\sP=\sa^\gamma\otimes\sa_\gamma$ and using  the identity $\sa^\beta_{,\alpha}\cdot\sa_\gamma=-\sa^\beta\cdot\sa_{\gamma,\alpha}=-\sa^\beta\cdot\sa_{\alpha,\gamma}$, which follows from \eqref{eq:196hh} and \eqref{eq:199g}, we find
\begin{equation}\label{eq:252bbb}
\begin{aligned}[b]
	(\sv_{,\beta}\cdot\sa^\alpha)\sP\sa^\beta_{,\alpha}&=(\sv_{,\beta}\cdot\sa^\alpha)(\sa^\gamma\otimes\sa_\gamma)\sa^\beta_{,\alpha}=(\sv_{,\beta}\cdot\sa^\alpha)(\sa^\beta_{,\alpha}\cdot\sa_\gamma)\sa^\gamma\\
	&=(\sv_{,\beta}\cdot\sa^\alpha)(\sa^\beta_{,\gamma}\cdot\sa_\alpha)\sa^\gamma=(\sv_{,\beta}\cdot(\sa^\alpha\otimes\sa_\alpha)\sa^\beta_{,\gamma})\sa^\gamma\\
	&=(\sv_{,\beta}\cdot\sP\sa^\beta_{,\gamma})\sa^\gamma=(\sv_{,\beta}\cdot\sP\sa^\beta_{,\alpha})\sa^\alpha\\
	&=(\sv_{,\beta}\cdot\sa^\beta_{,\alpha})\sa^\alpha-(\sv_{,\beta}\cdot(\bn\otimes\bn)\sa^\beta_{,\alpha})\sa^\alpha.
	\end{aligned}
\end{equation}
Moreover, since $\nabla\bn=\nabla\bn^\top$,
\begin{equation}\label{eq:253bbb}
\begin{aligned}[b]
	-(\sv_{,\beta}\cdot(\bn\otimes\bn)\sa^\beta_{,\alpha})\sa^\alpha&=	-(\bn\cdot\sv_{,\beta})(\bn\cdot\sa^\beta_{,\alpha})\sa^\alpha
	=	(\bn\cdot\sv_{,\beta})(\bn_{,\alpha}\cdot\sa^\beta)\sa^\alpha\\
	&=\nabla\bn^\top\nabla\sv^\top\bn=-\nabla\bn\nabla\bn\sv.
	\end{aligned}
\end{equation}
By combining \eqref{eq:251bbb}, \eqref{eq:252bbb}, and \eqref{eq:253bbb}, we obtain
\begin{equation}
\begin{aligned}[b]
\sP\operatorname{div}(\sP\nabla\sv^\top)-\nabla(\operatorname{div}\sv)&=-\nabla\bn^2\sv-(\operatorname{div}\bn)\nabla\sv^\top\bn\\
&=-\nabla\bn^2\sv+(\sP:\nabla\bn)\nabla\bn\sv.
\end{aligned}
\end{equation}
Recalling (\emph{cf.} \eqref{eq:227bbb} and \eqref{eq:228bbb}) that, for a spherical surface, $\nabla\bn=(1/R)\sP$ and $\operatorname{div}\bn=2/R$, we obtain,
\begin{equation}
\begin{aligned}
\sP\operatorname{div}(\sP\nabla\sv^\top)-\nabla(\operatorname{div}\sv)&=-\frac 1 {R^2}\sP\sv+\frac 2 {R^2}\sP\sv=\frac 1 {R^2}\sv,
\end{aligned}
\end{equation}
as required.
\paragraph{Proof of $\bf(\ref{eq:125})_{3,4,5}$}
First, we notice that
\begin{equation}
\sP\operatorname{div}\sP=-\sP\operatorname{div}(\bn\otimes\bn)=	-\sP((\operatorname{div}\bn)\bn+\nabla\bn\bn)=\mathbf 0.
\end{equation}
Then, using \eqref{eq:181bbb}, we obtain, for every scalar $\sigma$,
\begin{equation}
\sP\operatorname{div}(\sigma\sP)=\sP\nabla\sigma+\sigma\sP\operatorname{div}\sP=\sP\nabla\sigma.
\end{equation}
Now, $\eqref{eq:125}_{3,4}$ are arrived at by letting $\sigma=\operatorname{div}\sv$ and $\sigma=w/R$, respectively. Finally, to verify $\eqref{eq:125}_5$ we use again \eqref{eq:227bbb} to compute
\begin{equation}
\sP\operatorname{div}(\bn\otimes\nabla w)=\Delta w\sP\bn+\sP\nabla\bn	\nabla w=\frac 1 R \sP\nabla w=\frac 1 R\nabla w.
\end{equation}

\newpage

\end{document}